\documentclass[journal=jacsat,manuscript=article]{achemso}

\usepackage{graphicx}
\usepackage{dcolumn}
\usepackage{bm}

\usepackage[utf8]{inputenc}
\usepackage[T1]{fontenc}
\usepackage{float} 
\usepackage{mathptmx}
\usepackage{etoolbox}
\usepackage{color,soul}
\usepackage[version=4]{mhchem} 
\usepackage{comment}
\usepackage{geometry}
\usepackage{array}
\usepackage{makecell}

\usepackage{xr}
\makeatletter
\newcommand*{\addFileDependency}[1]{
  \typeout{(#1)}
  \@addtofilelist{#1}
  \IfFileExists{#1}{}{\typeout{No file #1.}}
}
\makeatother

\newcommand*{\myexternaldocument}[1]{%
    \externaldocument{#1}%
    \addFileDependency{#1.tex}%
    \addFileDependency{#1.aux}%
}

\myexternaldocument{supp_info}

\title[]{Physically motivated improvements of Variational Quantum Eigensolvers}

\author{Nonia Vaquero-Sabater}
 \affiliation{Donostia International Physics Center (DIPC), 20018 Donostia, Euskadi, Spain}
 \alsoaffiliation{Polimero eta Material Aurreratuak: Fisika, Kimika eta Teknologia Saila, Kimika Fakultatea, Euskal Herriko Unibertsitatea (UPV/EHU), PK 1072, 20080 Donostia, Euskadi, Spain.}

\author{Abel Carreras}
\affiliation{Multiverse Computing, 20014 Donostia, Euskadi, Spain}%

\author{Rom\'an Or\'us}
\affiliation{Donostia International Physics Center (DIPC), 20018 Donostia, Euskadi, Spain}
\alsoaffiliation{Multiverse Computing, 20014 Donostia, Euskadi, Spain}%
\alsoaffiliation{IKERBASQUE, Basque Foundation for Science, 48009 Bilbao, Euskadi, Spain}

\author{Nicholas J. Mayhall}
\affiliation{Department of Chemistry, Virginia Tech, Blacksburg, Virginia 24061, USA}

\author{David Casanova}
\affiliation{Donostia International Physics Center (DIPC), 20018 Donostia, Euskadi, Spain}
\alsoaffiliation{IKERBASQUE, Basque Foundation for Science, 48009 Bilbao, Euskadi, Spain}
\email{david.casanova@dipc.org}

\SectionNumbersOn

\begin{document}

\begin{abstract}
The Adaptive Derivative-Assembled Pseudo-Trotter Variational Quantum Eigensolver (ADAPT-VQE) has emerged as a pivotal promising approach for electronic structure challenges in quantum chemistry with noisy quantum devices.
Nevertheless, to surmount existing technological constraints, this study endeavors to enhance ADAPT-VQE's efficacy. Leveraging insights from electronic structure theory, we concentrate on optimizing state preparation without added computational burden and guiding ansatz expansion to yield more concise wavefunctions with expedited convergence toward exact solutions. These advancements culminate in shallower circuits and, as demonstrated, reduced measurement requirements. This research delineates these enhancements and assesses their performance across mono, di, and tridimensional arrangements of \ce{H4} models, as well as in the water molecule. Ultimately, this work attests to the viability of physically-motivated strategies in fortifying ADAPT-VQE's efficiency, marking a significant stride in quantum chemistry simulations.
\end{abstract}

\maketitle

\section{Introduction} \label{sec:intro}

Quantum computing stands at the forefront of a technological revolution poised to radically transform various fields, including quantum chemistry.\cite{Cao:quantum-chem:2019,McArdle:quantum-chem:2020} Its ability to harness the principles of quantum mechanics promises exponential computational speedup for certain classes of problems, offering a potential advantage over classical methods. 
Presently, quantum computers, characterized by a limited number of qubits and inherent operational imperfections, fall under the category of noisy intermediate scale quantum (NISQ) devices.\cite{Preskill2018quantumcomputingi,Brooks:quantum:2019} 
These systems, while powerful, are not yet capable of executing fault-tolerant quantum algorithms, necessitating the development of specific algorithms tailored for the NISQ era.
Within this paradigm, variational quantum eigensolvers (VQE)\cite{Peruzzo:2014,McClean_2016} have emerged as a prominent strategy for addressing electronic structure problems, particularly in the field of quantum chemistry.\cite{Kassal:quantum:2011,OMalley:quantum:2016,Tilly:vqe:2022}
VQE algorithms aim to approximate the ground state energy and wavefunction of the molecular electronic Hamiltonian. This is achieved through a variational approach, wherein a parameterized quantum circuit, known as the ansatz, is employed to encode trial wavefunctions. 
The energy of these wavefunctions is estimated through an optimization process combining both quantum and classical calculations. 
Typically, the quantum computer is responsible for evaluating the energy, while classical resources are used to optimize the parameters of the circuit.

In the context of quantum chemistry, wavefunction ansatz\"e used in VQE are typically based on the unitary coupled cluster (UCC) method.\cite{Bartlett:ucc:1989,Taube:ucc:2006} 
In this approach a reference wavefunction is expanded in the basis of molecular orbitals mapped to the qubit basis in a quantum circuit. 
Then, a parametrized unitary operator that encodes electronic transitions between these orbitals, is applied to the circuit to generate a particular trial wavefunction. 
This procedure allows to encode complex electronic states in a small number of qubits, i.e., in the order of the number of electrons. 
Notably, amongst the different variants of VQE, the adaptive derivative-assembled pseudo-Trotter VQE (ADAPT-VQE) approach\cite{Grimsley2019} has garnered attention for its adaptability and efficiency in constructing accurate wavefunctions.
ADAPT-VQE follows a dynamic strategy for selecting and refining the ansatz, in which the electronic wavefunction is systematically expanded to include those operators that contribute significantly to the energy.
ADAPT-VQE is particularly effective for strongly correlated electronic systems, where traditional classical methods may struggle. 
It has demonstrated significant promise in accurately predicting electronic ground state energies and properties, making it a valuable tool for quantum chemistry simulations.\cite{Tang:adapt:2021,Yordanov:adapt:2021,Grimsley:adapt:2023,Feniou:adapt:2023}
However, given present technological limitations, further modifications and variants of ADAPT-VQE are being explored to enhance its performance. 

In this study, we aim to optimize the efficiency of ADAPT-VQE by leveraging concepts from electronic structure theory in molecules. Specifically, our efforts address two crucial aspects of the algorithm by: (i) improving state preparation at the mean-field computational cost and (ii) guide the growth of the antsatz in order to produce more compact wavefunctions with faster convergence towards the exact solution. 
These improvements lead to shallower circuits and, as we demonstrate, fewer measurements.
The paper is organized as follows: section~\ref{sec:theory} introduces the main features of ADAPT-VQE and outlines our strategies to enhance its efficiency.
Section~\ref{sec:comput} describes the technical details employed in our simulations.
Then, we assess the performance of the new strategies in mono, di, and tridimensional arrangements of \ce{H4} models, as well as in the water molecule (section~\ref{sec:results}).
Finally, the key findings of our study are summarized in the Conclusions section.

\section{Theoretical background} \label{sec:theory}

\subsection{ADAPT-VQE}
The ADAPT-VQE ansatz is constructed through the sequential application of UCC-like exponentiated operators:
\begin{equation} \label{eq:vqe:ansatz}
    |\psi^{(N)}\rangle = \prod_{i=1}^N e^{\theta_i\hat{A}_i} |\psi^{(0)}\rangle, 
\end{equation}
where $|\psi^{(0)}\rangle$ denotes the initial state, and $\hat{A}_i$ represents the fermionic anti-Hermitian operator introduced during the $i$th iteration, with $\theta_i$ parameter denoting its corresponding amplitude.
A typical choice for the pool of accessible operators includes occupied-to-virtual or generalized (all-to-all) single and double excitations.
Beginning with $|\psi^{(0)}\rangle$, the ADAPT-VQE wavefunction grows iteratively by appending one exponentiated excitation operator to Eq.~\ref{eq:vqe:ansatz}. 
The new operator to be added at step $N+1$ is determined from the entire operator pool as the one yielding the largest gradient $\partial E^{(N)}/\partial\theta_i$, where:
\begin{equation}
    E^{(N)} = \langle\psi^{(N)}| \hat{H} |\psi^{(N)}\rangle.
\end{equation}
The set of parameters $\{\theta_i\}$ is optimized each time a new operator is introduced to the ansatz. 
Throughout this process, it is crucial to recycle the parameters $\{\theta_i\}$ between ADAPT iterations in order to circumvent undesirable local minima.\cite{Grimsley2023} 
Optimization of the amplitudes is performed on a classical computer, while the quantum circuit assesses energy and gradients. 

Despite the advancements of ADAPT-VQE in comparison to standard UCC-based VQEs\cite{Lee2019} in terms of accuracy and circuit depth,\cite{Grimsley2019,Tang2021,Yordanov2021} expanding the wavefunction based on the energy gradient does not guarantee convergence to the true ground state and is susceptible to becoming trapped in local minima of the potential energy surface.\cite{Grimsley2023}
In the following, we present physically-motivated simple strategies to improve the performance of Eq.~\ref{eq:vqe:ansatz} by addressing: (i) the shape of the initial state ($|\psi^{(0)}\rangle$), and (ii) the growth of the wavefunction, i.e., the selection of the excitation operators. 

\subsection{Improving initial state within mean-field} 

One of the critical factors influencing the success of ADAPT-VQE in efficiently capturing the true ground state electronic structure and energy lies in the fidelity of the initial state. The mean-field solution serves as an excellent initial approximation for the wavefunction of closed-shell molecules, characterized by significant energy gaps between the highest occupied molecular orbital (HOMO) and lowest unoccupied molecular orbital (LUMO). However, in strongly correlated systems, the overlap of the Hartree-Fock (HF) Slater determinant with the exact ground state diminishes, often constituting only 50\% or less of the exact non-relativistic solution, i.e., the full configuration interaction (FCI) expansion. It is precisely in these intriguing systems where traditional electronic structure methods encounter difficulties in describing the ground state, necessitating increasingly demanding computational resources to achieve chemical accuracy. 
Fortunately, these are the scenarios where quantum algorithms in quantum computers hold a distinct advantage over classical approaches.\cite{Cao:quantum-chem:2019}

To enhance the initial ground state estimate beyond the uncorrelated HF configuration, we envision utilizing the single-electron eigenstates of the one-particle density matrix (Eq.~\ref{eq:1pdm}), also known as the natural orbitals (NOs),\cite{Davidson:NO:1972} obtained from a computationally affordable correlated method,
\begin{equation} \label{eq:1pdm}
    \rho_{pq} = \langle\psi | \hat{a}_p^\dagger \hat{a}_q | \psi\rangle.
\end{equation}
In Eq.~\ref{eq:1pdm}, $\hat{a}_p^\dagger$ ($\hat{a}_q$) represents the creation (annihilation) operator for the $p$th ($q$th) spatial orbital. 
Here, we propose the use of density matrices from unrestricted Hartree-Fock (UHF). 
The UHF approach has the capability of reducing the computed energy by ``artificially'' breaking the spatial symmetry between $\alpha$-spin (spin-up) and $\beta$-spin (spin-down) orbitals. This symmetry breaking occurs in the presence of degeneracies or near-degeneracies at the Fermi level, while in closed-shell systems, UHF converges to the electron occupation of restricted orbitals following the Aufbau principle.\cite{Giesbertz:aufbau:2010}

Notably, the NOs from the UHF density permit fractional occupancies, mirroring features of correlated wavefunctions that go beyond mean-field theory, but with almost no additional computational cost with respect to (restricted) HF. 
Furthermore, the diagonalization of the total UHF density restores the spatial symmetry between the $\alpha$ and $\beta$ spin spaces (Figures~\ref{si:fig:h4_1d_1.5_orbitals} and~\ref{si:fig:h4_1d_3.5_orbitals}), potentially overcoming the limitations of symmetry broken UHF solutions as initial states in ADAPT-VQE.\cite{Bertels:adapt:2022}
In fact, the utilization of UHF NOs is a recognized, straightforward, and valuable strategy for selecting initial orbitals in multiconfiguration SCF wavefunctions, as employed in the complete active space SCF (CASSCF) approach.\cite{Pulay:UHF-NO}

The concept of enhancing orbitals within ADAPT-VQE has been recently exploited by Fitzpatrick et al., who employed an SCF orbital optimization strategy (ADAPT-VQE-SCF).\cite{fitzpatrick:adapt-scf:2022} However, it's important to note a distinction: while the incorporation of NOs in our approach focuses on enhancing the initial state, ADAPT-VQE-SCF continually updates orbitals as the ansatz expands, i.e., at each cycle of the ADAPT-VQE process.

\subsection{Guiding wavefunction growth with projection protocols}\label{projection_sub} 

Recently, Feniou and collaborators proposed an overlap-guided approach for a more efficient construction of ADAPT-VQE wavefunctions.\cite{Feniou2023} 
While the overlap criterion offers promise, it does require an accurate correlated reference wavefunction, which may be challenging to achieve at a moderate computational cost. 
On the other hand, the potential for generating more compact wavefunctions compared to the energy-gradient criterion suggests the feasibility of alternative strategies for optimizing the selection of excitation operators.

\begin{figure}[H]
    \centering
    \includegraphics[width=8cm]{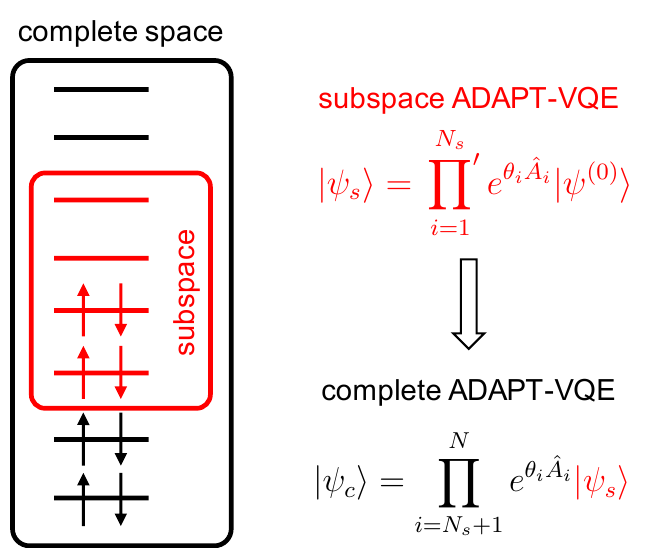}
    \caption{Left: Representation of the orbital subspace selection (in red). Right: ADAPT-VQE ansätze in the subspace ($s$, red) and complete ($c$, black) orbital spaces. The prime in the product of the $|\psi_s\rangle$ equation indicates that it only includes operators with orbital indices in the subspace.}
    \label{fig:orb_projection}
\end{figure}

To this end, we suggest a simpler criterion for enhancing the growth of ADAPT-VQE ansätze based on orbital energies. In general, without explicitly considering the spatial characteristics of molecular orbitals, the weight of excited configurations in the ground state wavefunction is inversely proportional to the energies of the involved orbitals, as outlined by second-order perturbation theory:
\begin{equation} \label{eq:mp2_ampl}
    t_{ij}^{ab} = - \frac{\langle ab || ij\rangle}{\varepsilon_a+\varepsilon_b-\varepsilon_i-\varepsilon_j},
\end{equation}
where $t_{ij}^{ab}$ represents the first-order M{\o}ller-Plesset (MP1) amplitude for the electronic configuration obtained by replacing occupied orbitals $i$ and $j$ with virtual orbitals $a$ and $b$, $\langle ab || ij\rangle$ denotes the two-electron antisymmetrized integral in physicist's notation, and $\varepsilon_p$ signifies the energy of the $p$th orbital.
Accordingly, it is reasonable to anticipate that excitation operators $\hat{A}_i$ involving molecular orbitals near the Fermi level — those with small denominators in Eq.~\ref{eq:mp2_ampl} — will play a significant role. Therefore, we initially restrict the orbital space to a subset of active orbitals, enabling a more cost-effective ADAPT-VQE. Subsequently, we project the resultant subspace ADAPT-VQE wavefunction onto the complete orbital space and resume the energy-gradient-driven ADAPT-VQE iterations until convergence, as depicted in Figure~\ref{fig:orb_projection}.

Besides the orbital energy criterion, the selection of the orbital subspace often relies on chemical intuition and prior knowledge of the molecule under study, akin to classical multiconfigurational approaches. Alternatively, it can be determined through active space selection schemes.\cite{sayfutyarova2017automated,bofill1989unrestricted,keller2015selection,bao2018automatic,khedkar2019active,sayfutyarova2019constructing,bao2019automatic,stein2016automated,veryazov2011select,Jeong:active:2020,King:active:2021}
In anycase, aside from the potential for generating compact ans\"atze by temporarily excluding a large number of typically less important operators from the pool, the initial subspace ADAPT-VQE also demands fewer computational resources, such as a smaller number of qubits.

We notice that the projection strategy depicted in Figure~\ref{fig:orb_projection} can be extended to 
use wavefunctions obtained with small basis sets as an effective starting point for calculations with larger basis, as in the dual-basis approximation,\cite{basis2,Liang:dual-basis:2004} 
which demonstrated considerable utility in enhancing the convergence and expediting (classical) electronic structure computations.
We have also implemented this approach in VQE through a basis transformation of the fermionic operators as:

\begin{equation} \label{eq:basis2}
a_p = \sum_i{a'_q \langle \chi'_q | \chi_p \rangle}
\end{equation}
where $a_p$ is a generic operator that acts on molecular orbital $p$ of the target basis, $\langle \chi'_q | \chi_p \rangle$ is the overlap integral between the molecular orbitals of the original $\{\chi'_q\}$ and target $\{\chi_p\}$ basis, and $a'_q$ are operators defined in the original basis.
While this strategy may be easily exploitable in standard VQE, we anticipate its effectiveness to be less pronounced in ADAPT-VQE. This is because the intrinsic Trotterized shape of the ADAPT-VQE wave function leads to the construction of large ans\"atze. However, it might still prove to be a viable strategy for ADAPT-VQE by truncating Eq.~\ref{eq:basis2} for large values of the overlap integral, thereby significantly reducing the number of terms in the ansatz.

\section{Computational details} \label{sec:comput}

All simulations have been done with a custom implementation of ADAPT-VQE in Python using NumPy,\cite{numpy} SciPy,\cite{2020SciPy} and OpenFermion\cite{mcclean2019openfermion} packages.
The mapping between fermionic and spin operators has been done following the Jordan-Wigner (JW) transformation.\cite{JW:mapping:1928}
Assessments of the Hamiltonian in sections~\ref{sec:h4} and~\ref{sec:h2o} have been done using sparse matrices. 
The building and measurement of quantum circuits in section~\ref{sec:depth} have been performed with the Qiskit platform.\cite{Qiskit}
In all cases, the energy has been evaluated without including quantum noise. 
Numerical optimizations of amplitudes $\{\theta_i\}$ have been performed with the Broyden–Fletcher–Goldfarb–Shanno (BFGS)\cite{BFGS} algorithm as implemented in the SciPy module, except for the measurement of circuit depths (section~\ref{sec:depth}), where the constrained optimization by linear approximation (COBYLA) optimizer\cite{Powell:cobyla:1994} was employed.
The convergence criteria for ADAPT-VQE is controlled through the tolerance value $\delta$ for the classical optimizer. 
Convergence is achieved when the energy for an ADAPT-VQE iteration is above the previous one. We set $\delta=10^{-4}$ Hatrees for all simulations.
The choice of this rather large threshold is motivated by the practical limitations of quantum hardware to reach energy errors below $10^{-4}$ Hartrees due to statistical noise.

In all cases, we have employed operator pools composed by UCC operators restricted to occupied-to-virtual spin-singlet adapted single and double excitations.\cite{Grimsley2019} 
Notice that the combined use of restricted orbitals (same spatial orbitals for $\alpha$ and $\beta$ spins) and spin-adapted operators ensure spin completeness of the ans\"atze.
Therefore, all the results in the present study correspond to pure spin singlet states (no spin contamination, $\langle \hat{S}^2\rangle=0$).
It is important to acknowledge that the direct implementation of spin-adapted fermionic operators on quantum hardware poses significant technical challenges due to the fact that the individual terms don’t commute, and thus don’t factorize into a simple sequence of pauli rotations. However, because our current goal is to explore the ability of NO’s to accelerate the recovery of electron correlation with ADAPT-VQE, we wanted to remove the possibility that spin-contamination could be allowing the state to converge to a broken-symmetry solution. Such a spin-contaminated solution could be interpreted as a false positive, where at longer bond-lengths, the ADAPT-VQE energies might be accurate while the state fidelities would remain poor. However, we don’t expect the qualitative conclusions to change if a more hardware efficient pool were to be used.
The necessary one and two-electron integrals used to construct the molecular Hamiltonian have been computed with PySCF.\cite{pyscf} 
As we aim to evaluate the ability of various ADAPT-VQE flavors to recover electron correlation effects beyond the minimal basis set, we perform all the calculations with the 3-21G basis.
The projection between partial and complete orbital spaces has been done within the same basis set, as indicated in section~\ref{sec:results} for each case study.
Unless indicated, calculations for the \ce{H4} models have been done with all orbitals being active (8 orbitals), while we consider 10 active orbitals in the water molecule simulations, that is, 3 inactive orbitals (the oxygen's 1s and the two highest virtual orbitals).
Subspace calculations were performed with half of the active space, i.e., 4 (\ce{H4} models) and 6 (water molecule) orbitals.

\section{Results and discussion} \label{sec:results}

In the following, we evaluate the performance of the two introduced approaches, i.e., enhancing the fidelity of the initial state using UHF NOs and expanding the ADAPT-VQE ansatz through the subspace-to-complete wavefunction projection. To do so, we consider the \ce{H4} model with three spatial configurations: linear (1-dimensional), square (2-dimensional), and tetrahedral (3-dimensional), each with \ce{H-H} distances of 1.5 and 3.0~{\AA}. Additionally, we assess the effectiveness of our methods in computing the electronic energy of the water molecule with a \ce{O-H} bond length of 1.0~{\AA} (near equilibrium), as well as with both \ce{O-H} bonds stretched at 3.0~{\AA}. 
The \ce{HOH} molecular angle was fixed to the experimental value at equilibrium (104.5$^\circ$).\cite{HOY19791}

\subsection{Applications to \ce{H4} systems}\label{sec:h4}

\subsubsection{Energy convergence}

\begin{figure}[H]
    \centering
    \includegraphics[width=6.5cm]{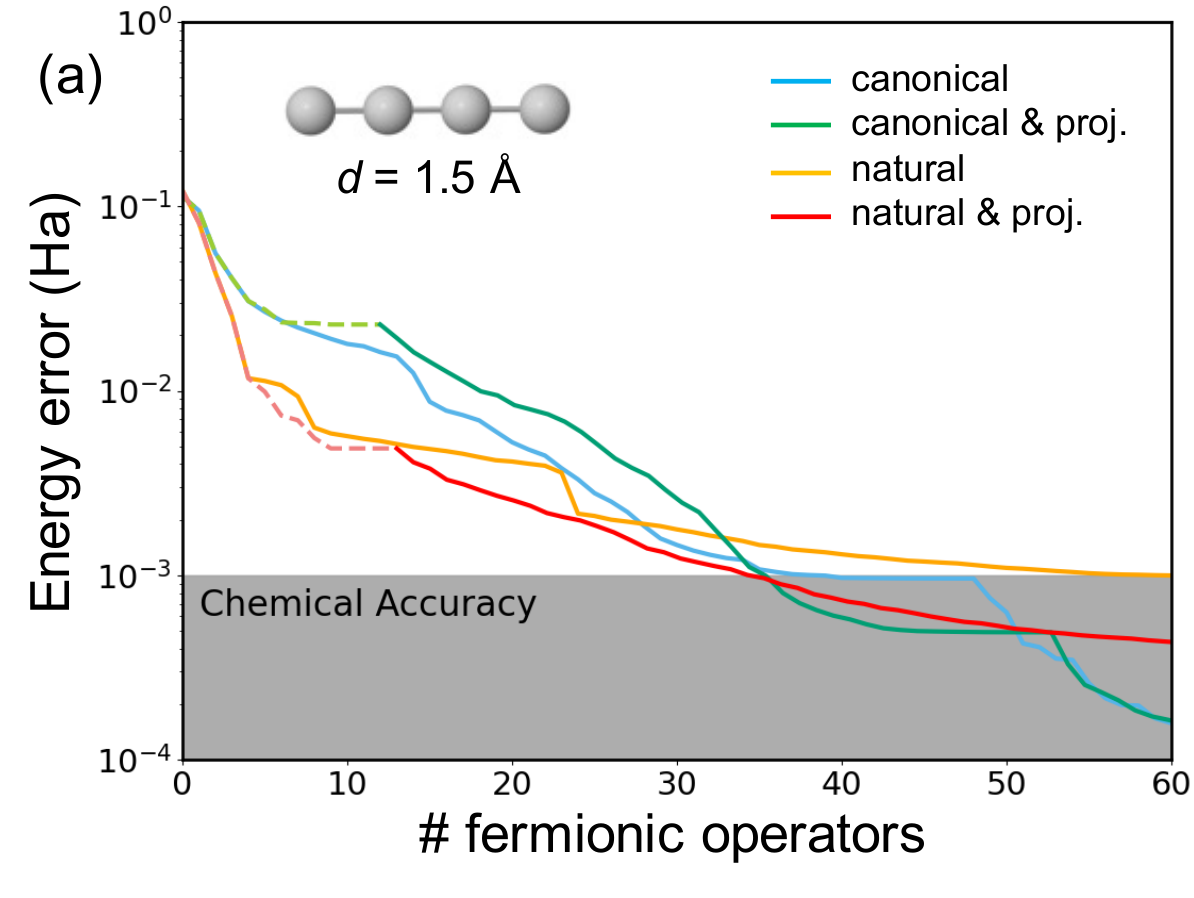}
    \includegraphics[width=6.5cm]{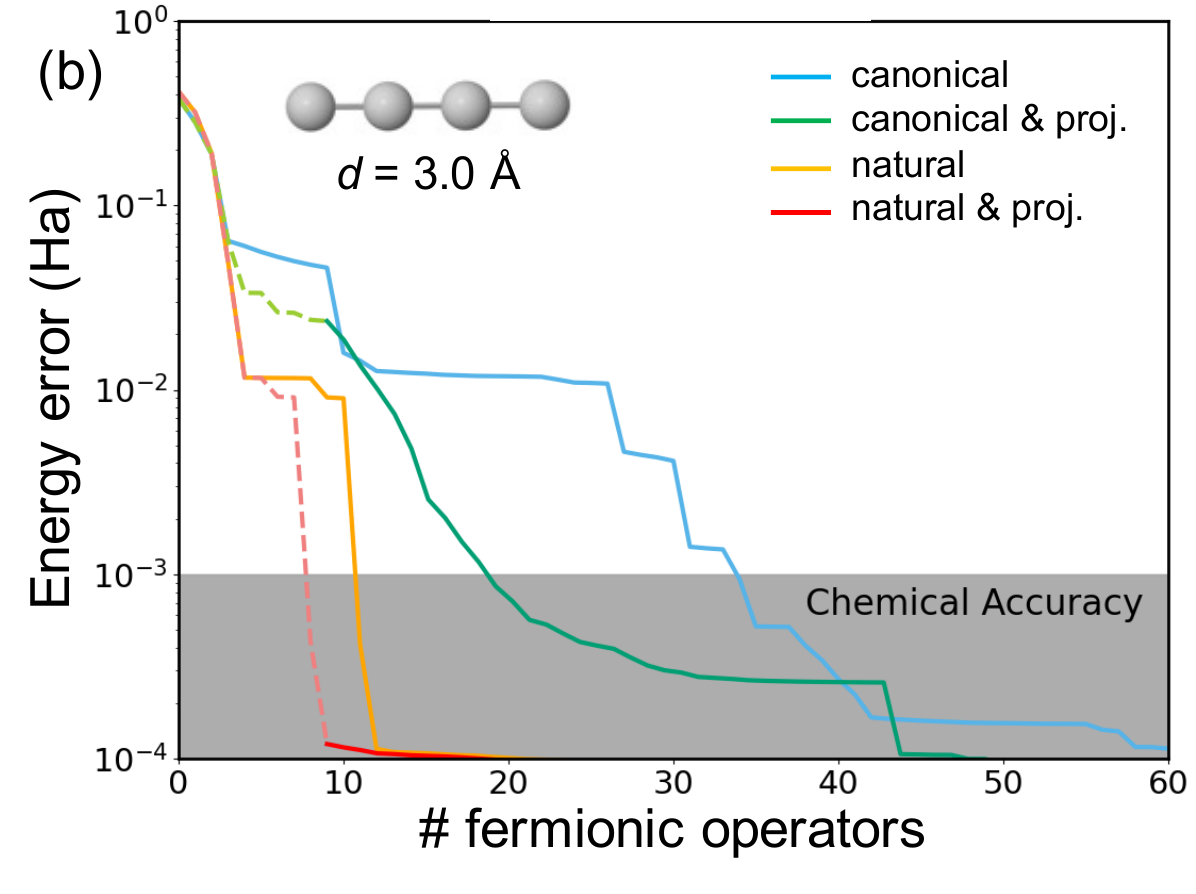}
    \includegraphics[width=6.5cm]{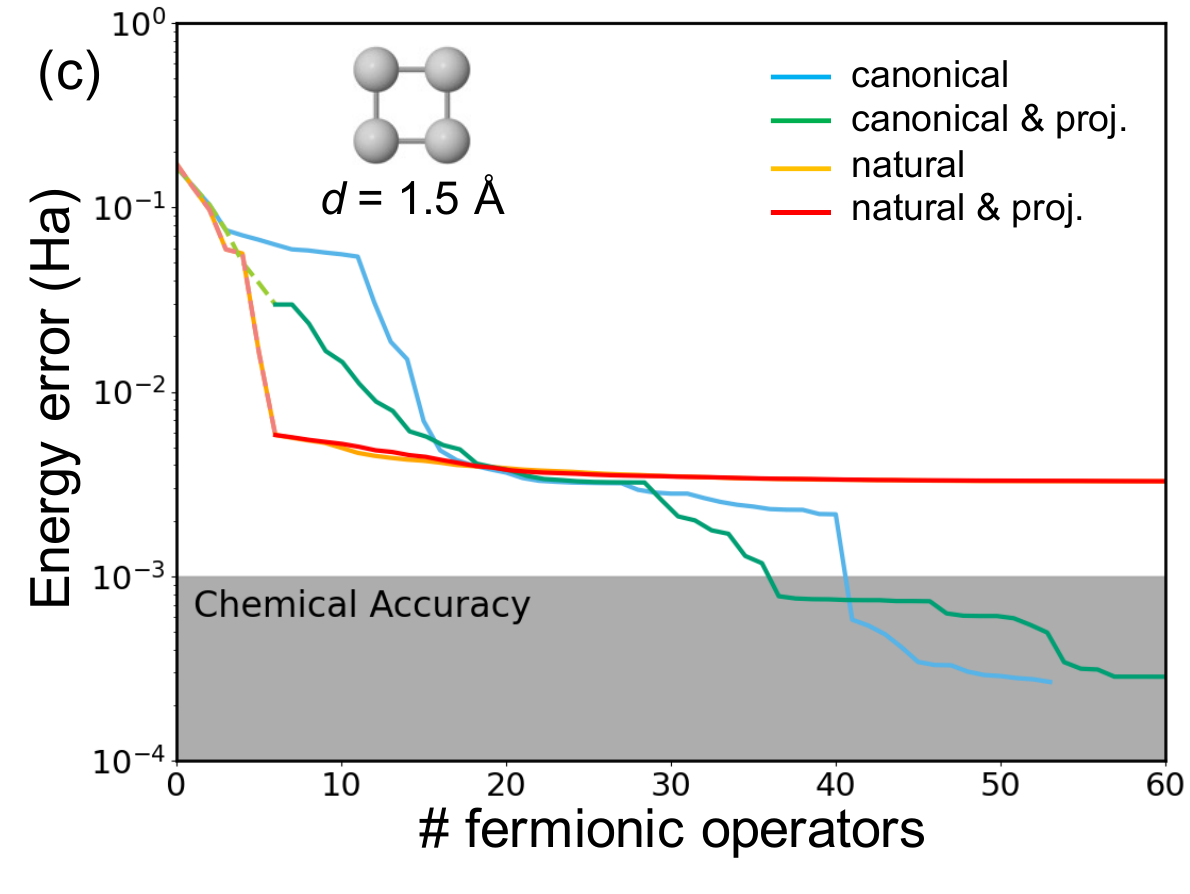}
    \includegraphics[width=6.5cm]{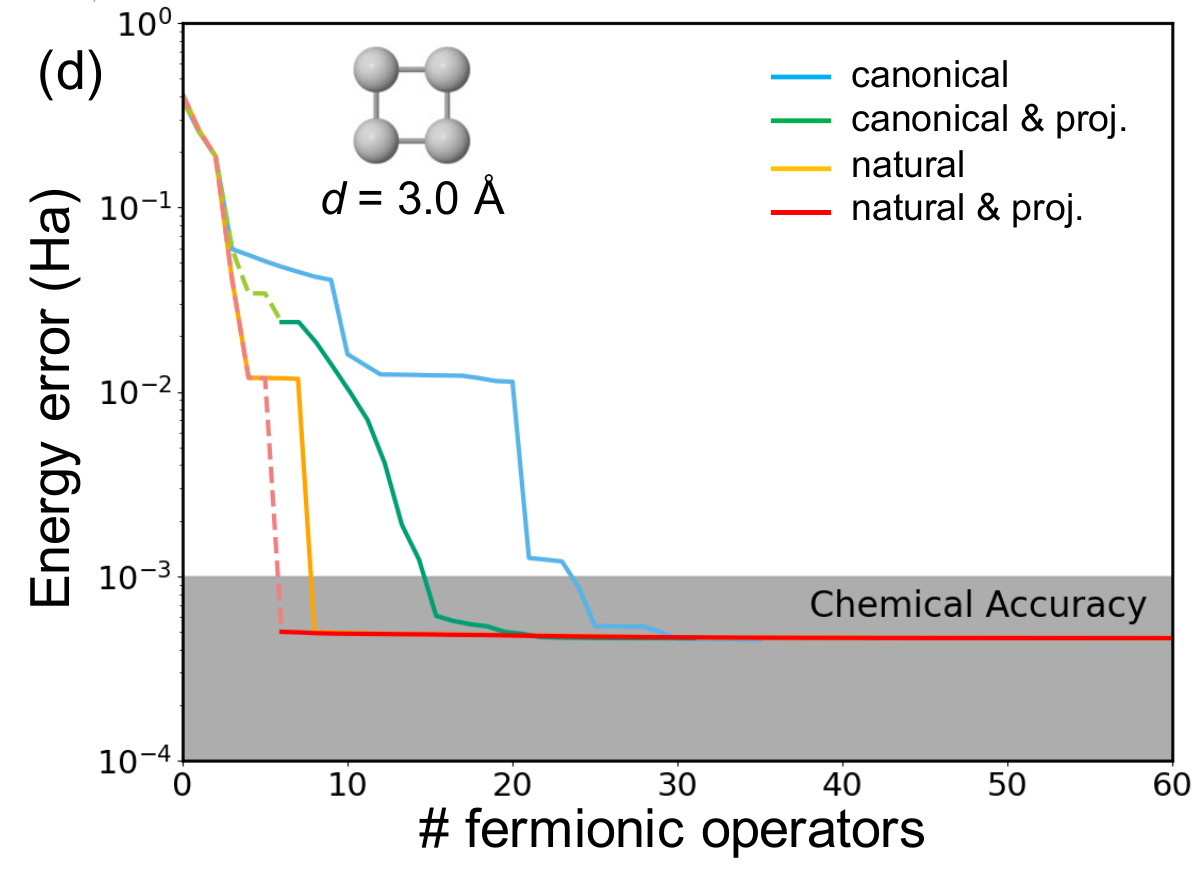}
    \includegraphics[width=6.5cm]{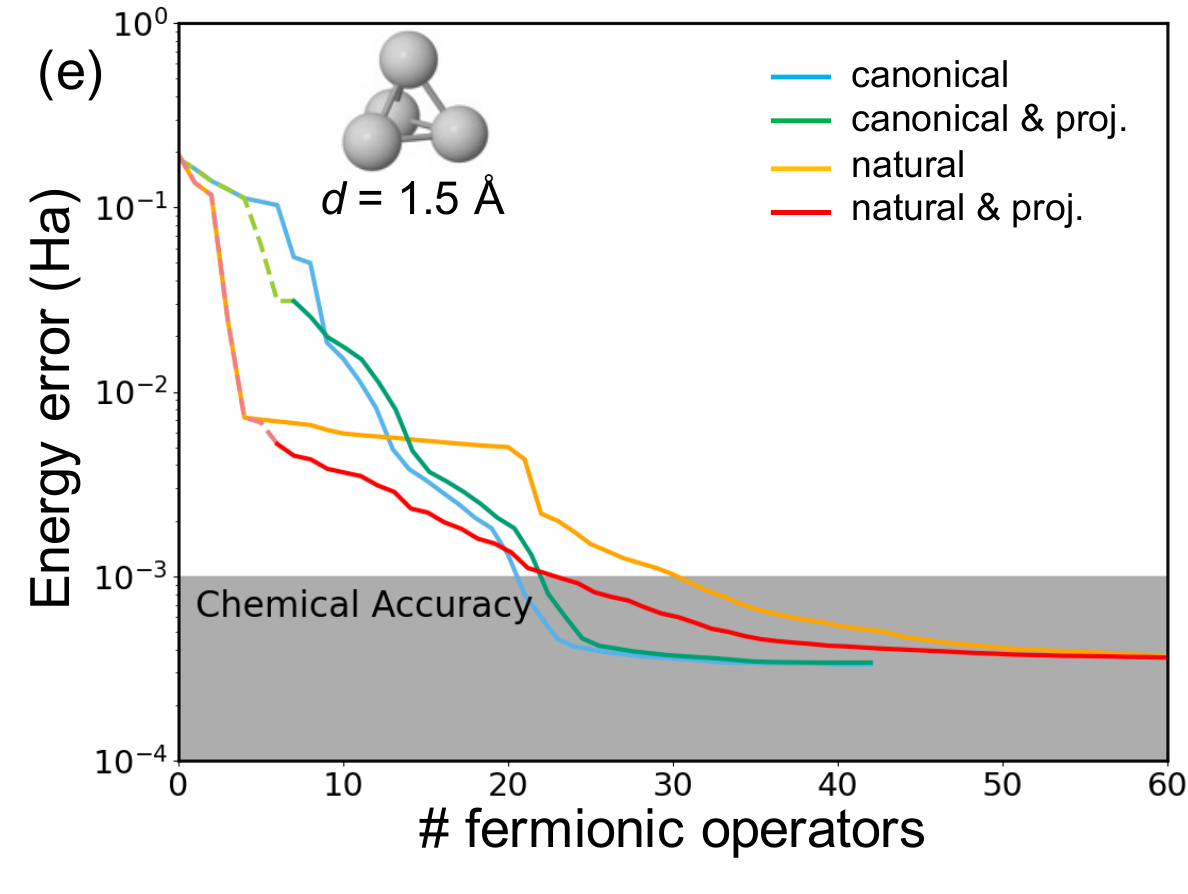}
    \includegraphics[width=6.5cm]{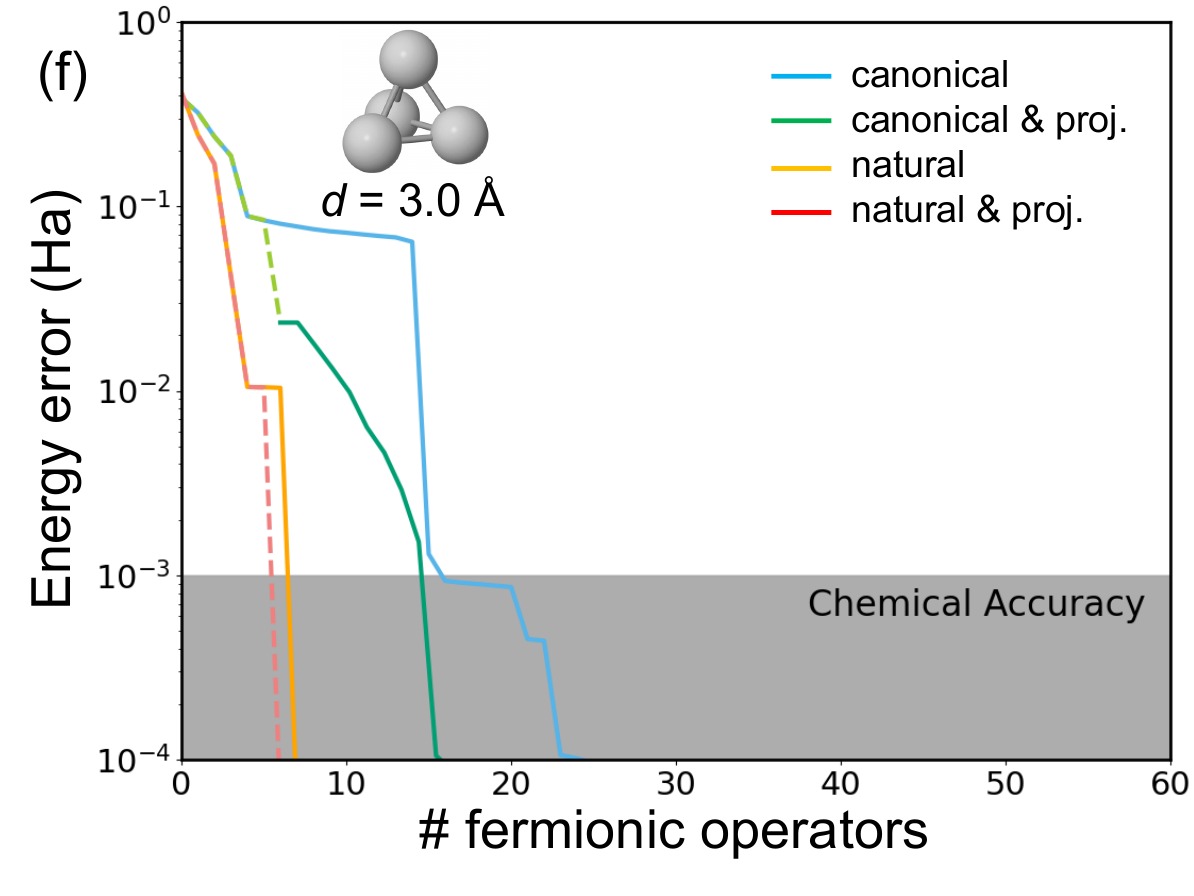}
    \caption{Energy error (in Hartree) of ADAPT-VQE wavefunctions with respect to FCI as a function of the number of fermionic operators obtained for the linear (a, b), square (c, d) and tetrahedral (e, f) \ce{H4} systems with \ce{H-H} distance of 1.5~{\AA} (a, c, and e) and 3.0~{\AA} (b, d, and f), performed with canonical (blue), projected canonical (green), natural (orange), and projected natural (red) orbitals.
    Initial iterations with orbital subspaces indicated with dashed lines.
    }
    \label{fig:h4_energies}
\end{figure}

The energy convergence of \ce{H4} models with respect to the number of fermionic operators included in the ansatz, which is directly related to the depth of the circuit, is quite sensible to the choice of the initial state and the use of the orbital projection strategy (Figure~\ref{fig:h4_energies}).

In general, the ADAPT-VQE energy errors, defined as deviations with respect to the FCI solution, exhibit an initial accelerated decay for the explored 1D, 2D, and 3D \ce{H4} models when employing NOs. 
The advantage of utilizing NOs, as opposed to the canonical MOs, to represent the initial state becomes notably pronounced at larger interatomic separations ($d = 3$ {\AA}). 
At this distance, the number of operators required to achieve chemical accuracy is reduced by a factor of 2 to 4 compared to those obtained with mean-field orbitals. 
This reduction implies significantly shallower quantum circuits.
At the shorter $d = 1.5$ {\AA} \ce{H-H} distance, improvements are notably more modest, displaying performances akin to those achieved with canonical MOs. 
We note that ADAPT-VQEs constructed with NOs exhibit an initial rapid reduction in the energy error. However, as more fermionic operators are incorporated into the wavefunction, the advantage over mean-field orbitals becomes less conspicuous. 
These findings suggest that, at short distances, achieving chemical accuracy may necessitate approximately the same circuit depth when employing both natural and mean-field orbitals. 
The square configuration (Figure~\ref{fig:h4_energies}c) stands as an exception, as despite the initial faster energy minimization encountered with NOs, it eventually exhibits poorer performance than with canonical orbitals.

Orbital projection from an initial subspace of 4 MOs ($N_s=4$) to the complete orbital set ($N=8$) enhances the performance of conventional ADAPT-VQE for $d = 3.0$~{\AA}, although not to the extent observed with NOs.
At $d = 1.5$~{\AA}, the energy error decay profiles closely resemble those obtained with the standard approach. 
However, as discussed in subsection~\ref{projection_sub}, employing an orbital subspace in the initial set of ADAPT cycles (indicated by dashed lines in Figure~\ref{fig:h4_energies}) entails a reduced computational cost.

By combining both strategies — utilizing the NO basis and incorporating the orbital projection scheme — we observe slightly improved performances across all configurations and distances compared to using NOs alone (without orbital projection). 
Additionally, this hybrid approach offers the advantage of conducting a portion of the simulation with a reduced number of qubits, inherent to the projection scheme.

These findings suggest that the utilization of NOs and the orbital projection strategy are particularly efficient in describing (low-spin) ground state molecules with unpaired electrons, indicative of strongly correlated systems.
Such systems pose a significant challenge for classical quantum chemistry approaches, necessitating high-hierarchy (multiconfigurational) electronic wavefunctions. This efficiency can be attributed to the decreasing overlap between the RHF solution and exact ground state wavefunction as the state acquires a more multiconfigurational character.
The symmetrized UHF based strategy, facilitated by the use of its NOs, offers a notably improved scheme. 
This improvement is attributed to its ability to partially introduce the effect of fractional orbital occupancies, enhancing the description of states with significant multiconfigurational character. 
Furthermore, electronic configurations beyond the HF-like state, carrying substantial weight in the ground state wavefunction, involve single or multiple electronic promotions between the occupied-virtual frontier orbitals. 
Notably, these configurations align with the terms present in the orbital subspace operator pool employed in the orbital projection procedure. 
In other words, the orbital subspace restriction facilitates the inclusion of these contributions right from the initiation of the ansatz growth.

\subsubsection{Fidelity of the ans\"atze}

\begin{figure}[H]
    \centering
    \includegraphics[width=6.5cm]{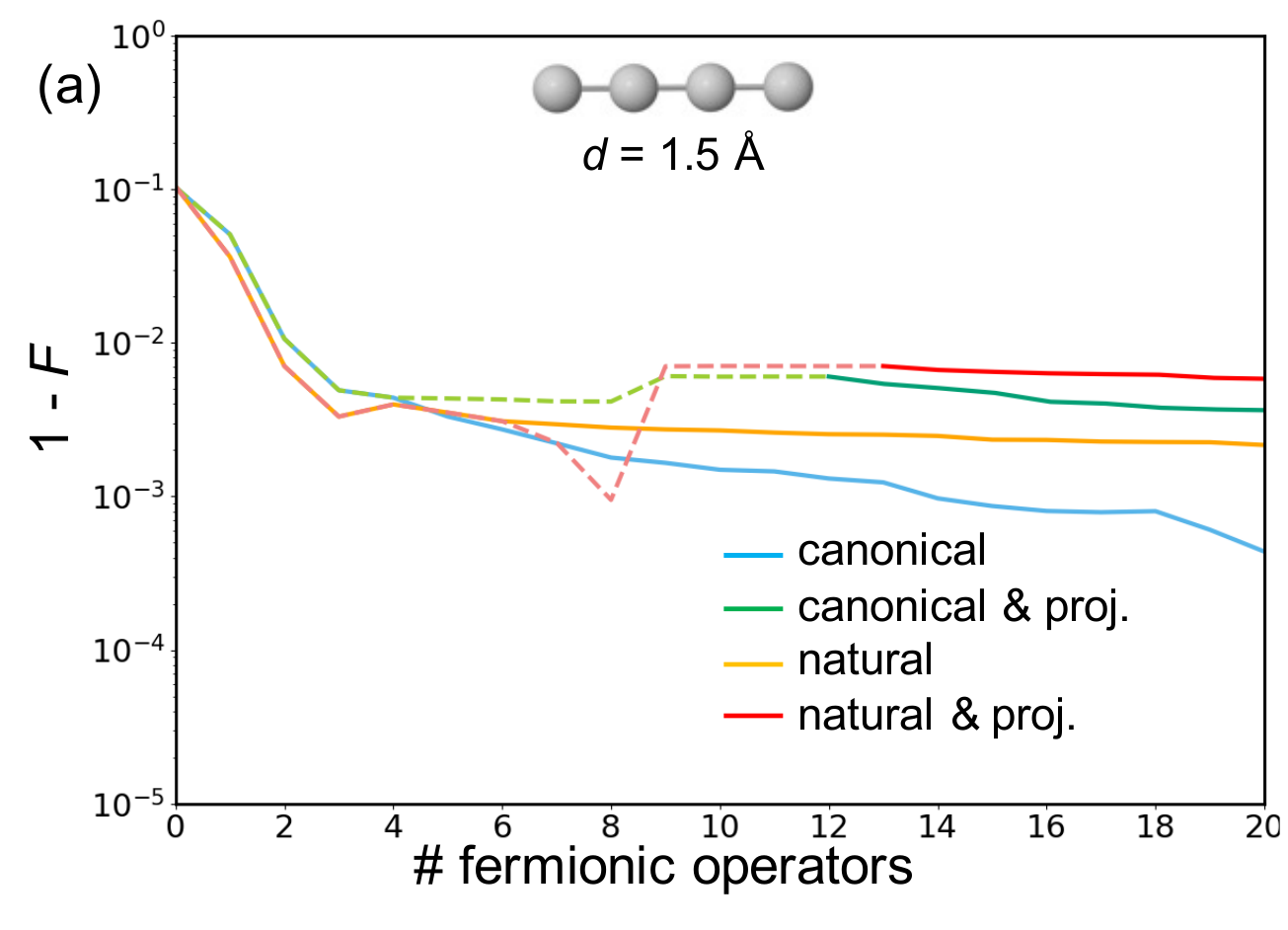}
    \includegraphics[width=6.5cm]{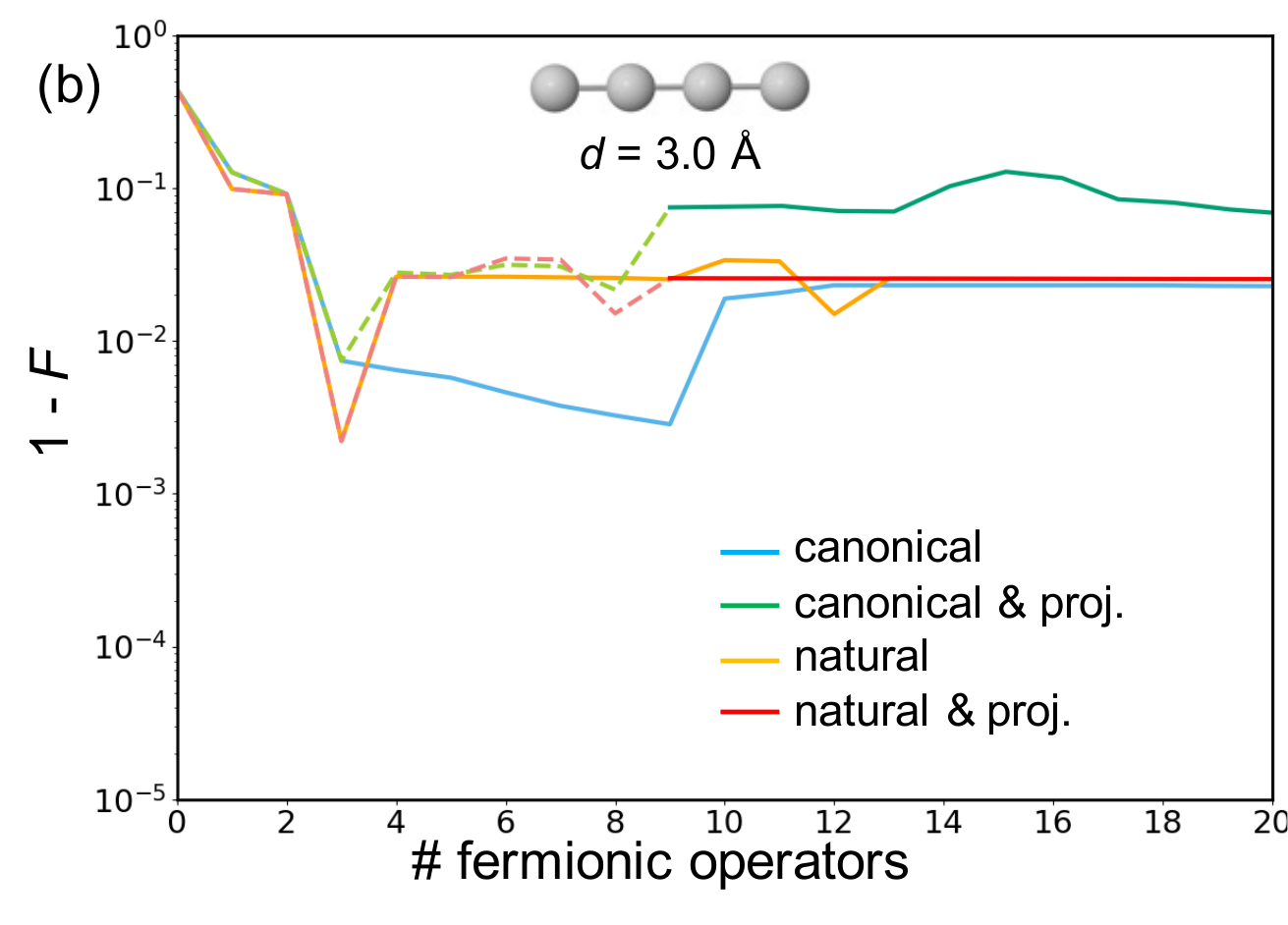}
    \caption{Fidelity of ADAPT-VQE wavefunctions with respect to FCI obtained for the linear \ce{H4} system with \ce{H-H} distance of 1.5~{\AA} (a) and 3.0~{\AA} (b), performed with canonical (blue), projected canonical (green), natural (orange), and projected natural (red) orbitals.
    Initial iterations with orbital subspaces indicated with dashed lines.}
    \label{fig:h4_fidelities}
\end{figure}

In the following, we aim to evaluate the quality of the obtained ADAPT-VQE ans\"atze.
For that, we employ the fidelity measurement\cite{Jozsa:fidelity:1994} of the density matrix obtained at each step of the performed simulations, defined as:
\begin{equation} \label{eq:fidelity}
    F(\rho_{ref}, \rho) = \left(tr  \sqrt{\sqrt{\rho_{ref}} \rho \sqrt{\rho_{ref}} } \right)^2,
\end{equation}
where $\rho_{ref}$ and $\rho$ are the density matrices of the exact (FCI) and estimated wavefunctions, respectively. 
The $F$ function is a measure of distance between density operators that takes values between 0 and 1, where 1 indicates identical density matrices. 
For pure states, as those explored here, $F$ reduces to the projection of the trial wavefunction with the reference solution, $F(\rho_{ref}, \rho) = |\langle\Psi | \Psi_{ref}\rangle|^2$.
Figure~\ref{fig:h4_fidelities} represents the evolution of $(1 - F)$ in terms of the number of operators in the ansatz for the different flavors of ADAPT-VQE seen in the previous section. Results for the square and tetrahedral systems can be found in the Supporting Information (Figures~\ref{si:fig:h4_square_fidelity} and~\ref{si:fig:h4_td_fidelity}).

\begin{figure}[H]
    \centering
    \includegraphics[width=6.5cm]{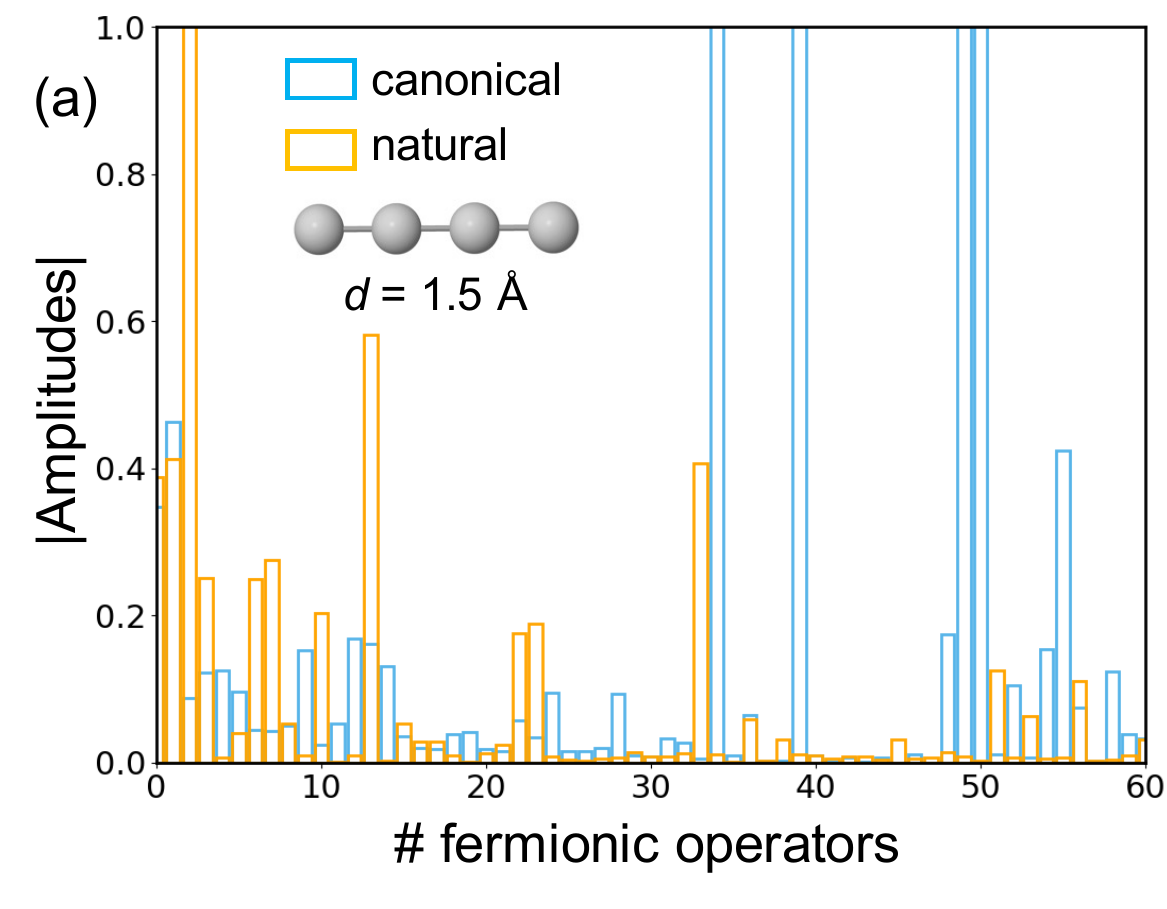}
    \includegraphics[width=6.5cm]{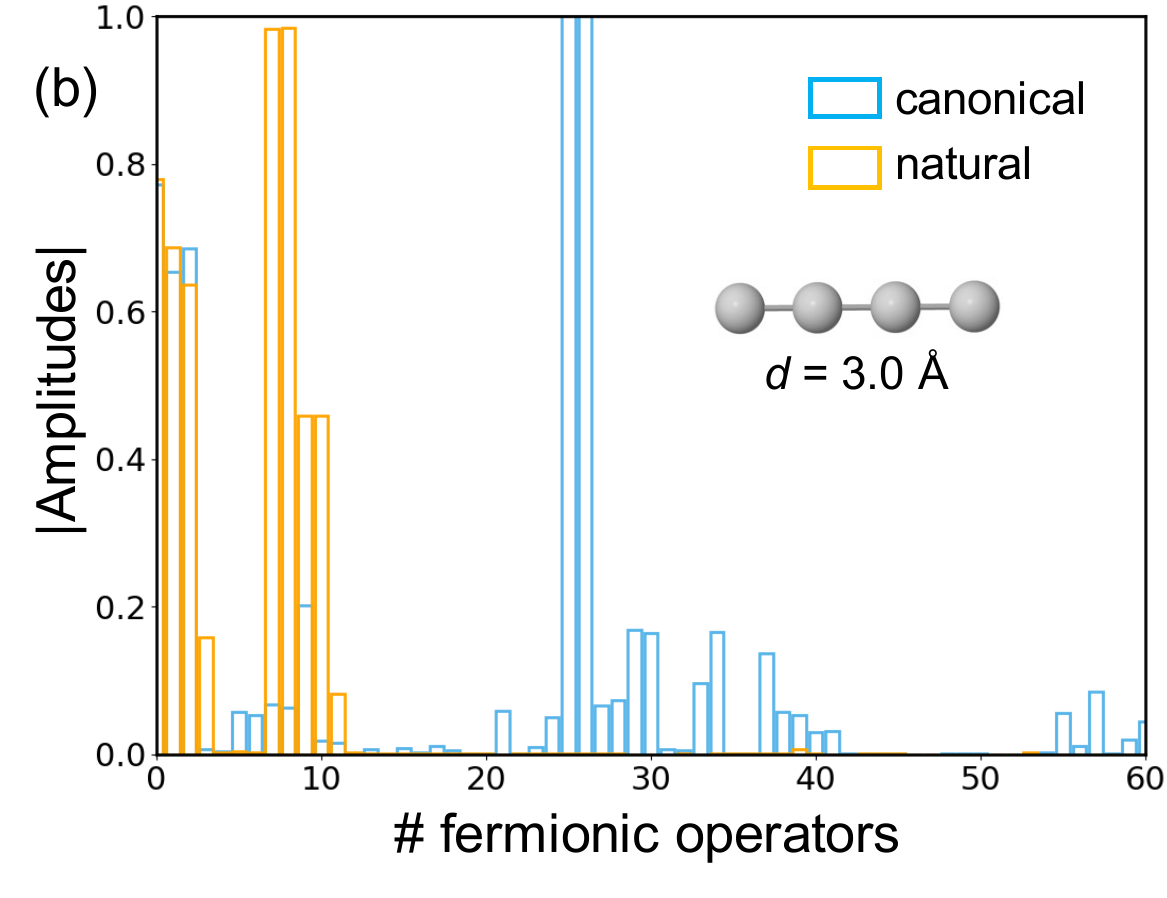}
    \caption{Absolute amplitudes $\{|\theta_i|\}$ of the fermionic operators in the final ansatz of the linear \ce{H4} system with \ce{H-H} distance of 1.5~{\AA} (a) and 3.0~{\AA} (b), performed with canonical (blue) and natural (orange) orbitals.}
    \label{fig:h4_amplitudes}
\end{figure}

For all \ce{H4} arrangements, we observe a similar fast initial decay of $(1-F)$, almost independently of the employed strategy.
This behaviour is concomitant with the decrease in the energy errors (Figure~\ref{fig:h4_energies}), which is a testimony of the large impact of the first operators added. 
On the other hand, after the few initial operators, fidelities seem to reach a plateau.
Moreover, ans\"antze with larger fidelities do not necessarily have smaller energy errors, as seen, for instance, in the linear \ce{H4} at $R=3$ {\AA} for wavefunctions with $\sim20$ operators.
In other words, while energy and fidelity follow similar trends initially, after a few ADAPT steps the decay in energy errors does not translate to higher fidelities (Figure \ref{fig:h4_amplitudes}).

This lack of correlation between energy errors and fidelity of ans\"atze approaching chemical accuracy seems to indicate that gradient-driven ADAPT-VQEs are rather efficient in order to incorporate those excitation operators with the largest impact on the molecular energies, even though these are not the ones providing the largest fidelities according to Eq.~\ref{eq:fidelity}. 
This can be particularly magnified at long bond distances, where there exists another singlet state close in energy to the ground state that may hinder the convergence of the VQE to the true ground state (Figure {\ref{si:fig:h4_fullci_states}}).
Therefore, we conclude that despite that ideally the ansatz in ADAPT-VQE should converge to the exact wavefunction (and density matrix), the lack of correlation between the errors in the molecular energy and the fidelity measurement seems to advice against the use of the latter to guide the growth of ADAPT-VQE solutions to the molecular Hamiltonian with targeted low-energy errors, e.g., chemical accuracy.

Despite the similarities observed in the fidelity profiles generated by canonical and natural orbitals, a notable distinction emerges in the structural characteristics of the two ans\"atze. 
While the fermionic amplitudes optimized using canonical orbitals display a dispersed distribution, with significant excitation terms often not manifesting until later iterations (Figure~\ref{fig:h4_amplitudes}), those derived from NOs yield more compact wavefunctions, in which the initially introduced terms hold the largest weights.
These findings elucidate the efficacy of NOs in achieving improved energy outcomes while employing a reduced number of operators. The disparities in the amplitude distributions between MOs and NOs are particularly conspicuous at large \ce{H-H} distances ($d=3.0$~\AA), aligning with the observed accelerated convergence towards chemical accuracy (Figure~\ref{fig:h4_energies}b).

\subsubsection{Size of the active space} 

\begin{figure}[H]
    \centering
    \includegraphics[width=6.8cm]{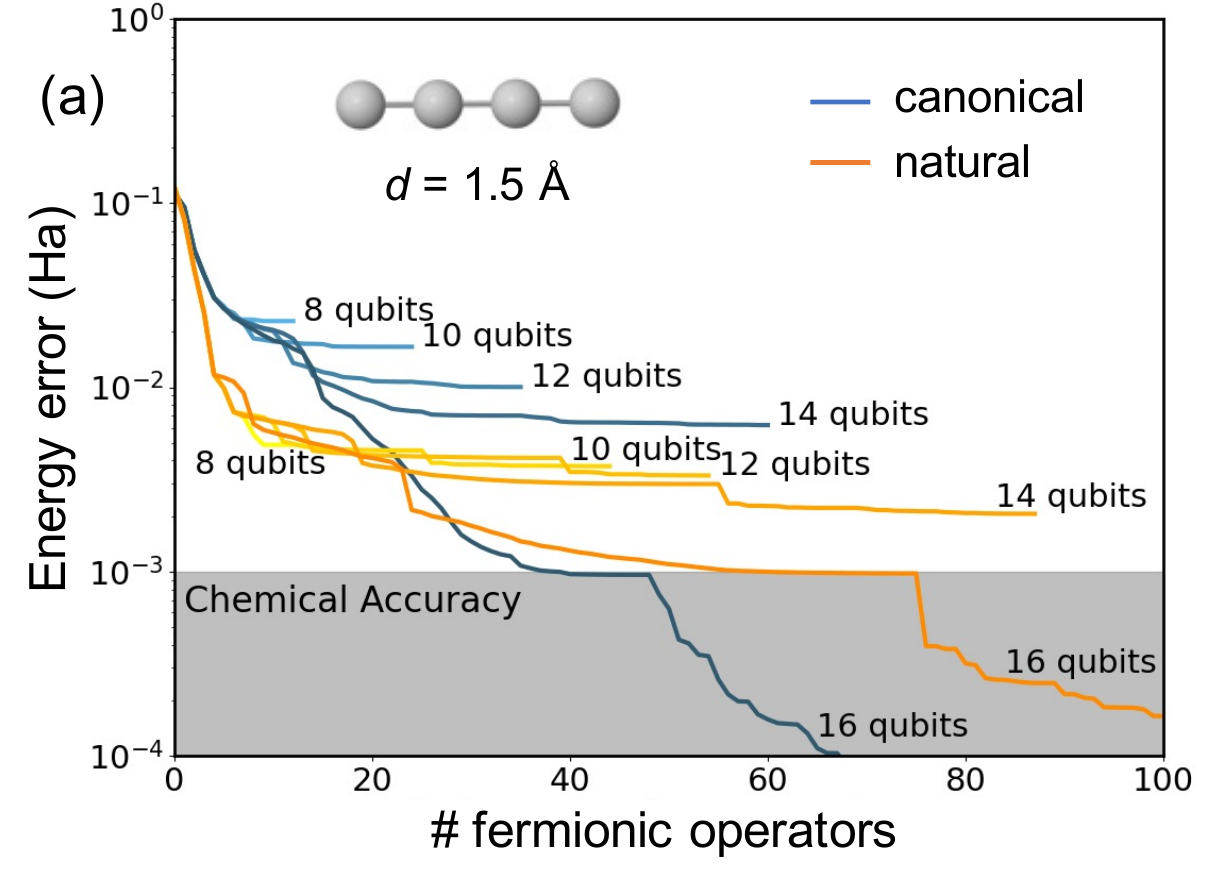}
    \includegraphics[width=6.8cm]{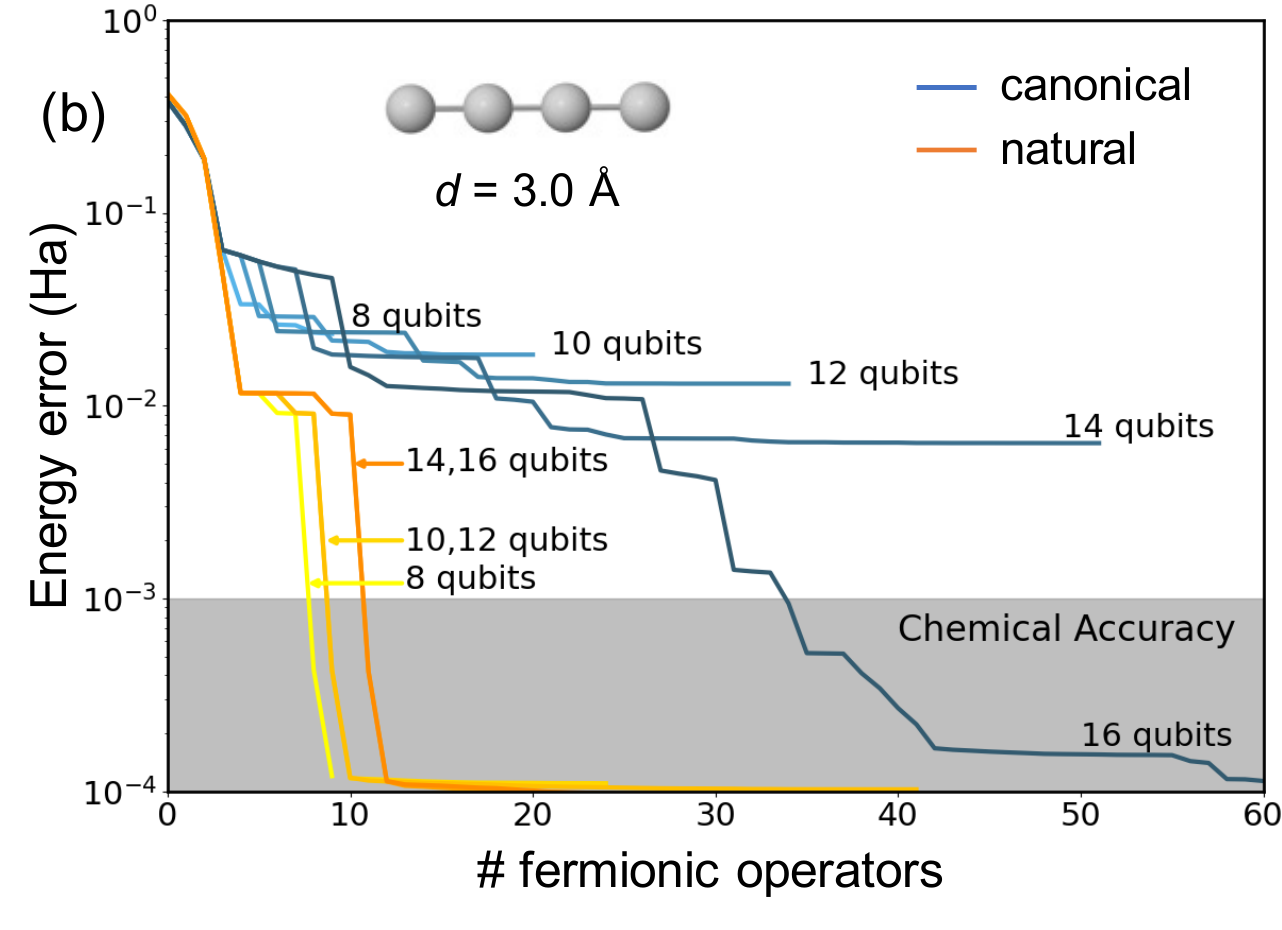}
    \caption{Energy errors (in Hartree) with respect to FCI obtained with different number of active orbitals, i.e., qubits (from 8 to 16), for the linear \ce{H4} system with 1.5~{\AA} (a) and 3.0~{\AA} (b) \ce{H-H} distances performed with canonical (light-dark blue) and natural (yellow-orange) orbitals.
    }
    \label{fig:h4_linear_orbs}
\end{figure}

To deepen our comprehension of the dependence of ADAPT-VQE electronic energies with the size of the orbital space in the MO and NO basis, we conduct a detailed analysis of energy errors across varying numbers of active orbitals. 
To this end, we focus on the linear \ce{H4} system with interatomic distances of 1.5 and 3.0~{\AA}. 
Comparable results for the square and pyramidal arrangements are presented in the Supporting Information (Figures~\ref{si:fig:h4_square_act} and~\ref{si:fig:h4_td_act}).
Figure~\ref{fig:h4_linear_orbs} shows ADAPT-VQE energy errors across active spaces ranging from 4 to 8 orbitals, corresponding to 8 to 16 qubits.

Expanding the active space size is anticipated to yield smaller errors upon convergence, as increased access to orbitals enhances the ansatz's expressibility. Conversely, the utilization of restricted orbital spaces may positively influence the wavefunction's shape, resulting in lower energies during the initial ADAPT cycles. This phenomenon is evident in the stretched chain ($d=3.0$~{\AA}), where employing smaller active spaces guides the ansatz's growth, leading to faster convergence toward the exact solution.
These results further justify the use of orbital projection schemes, particularly in open-shell systems.
Notably, energy convergence to chemical accuracy is significantly swifter with NOs, outperforming canonical MOs in all cases.
The lone exception is the scenario with $d = 1.5$~{\AA} using 16 qubits (Figure~\ref{fig:h4_linear_orbs}a). However, even in this case, the use of NOs results in smaller errors during the initial iterations (with a small number of fermionic operators).
In summary, these findings underscore the advantage of using NOs when dealing with reduced active orbital spaces, showcasing a reduction in errors by at least one order of magnitude.


\subsubsection{Energy measurements in the classical optimizer}

\begin{figure}[H]
    \centering
    \includegraphics[width=7cm]{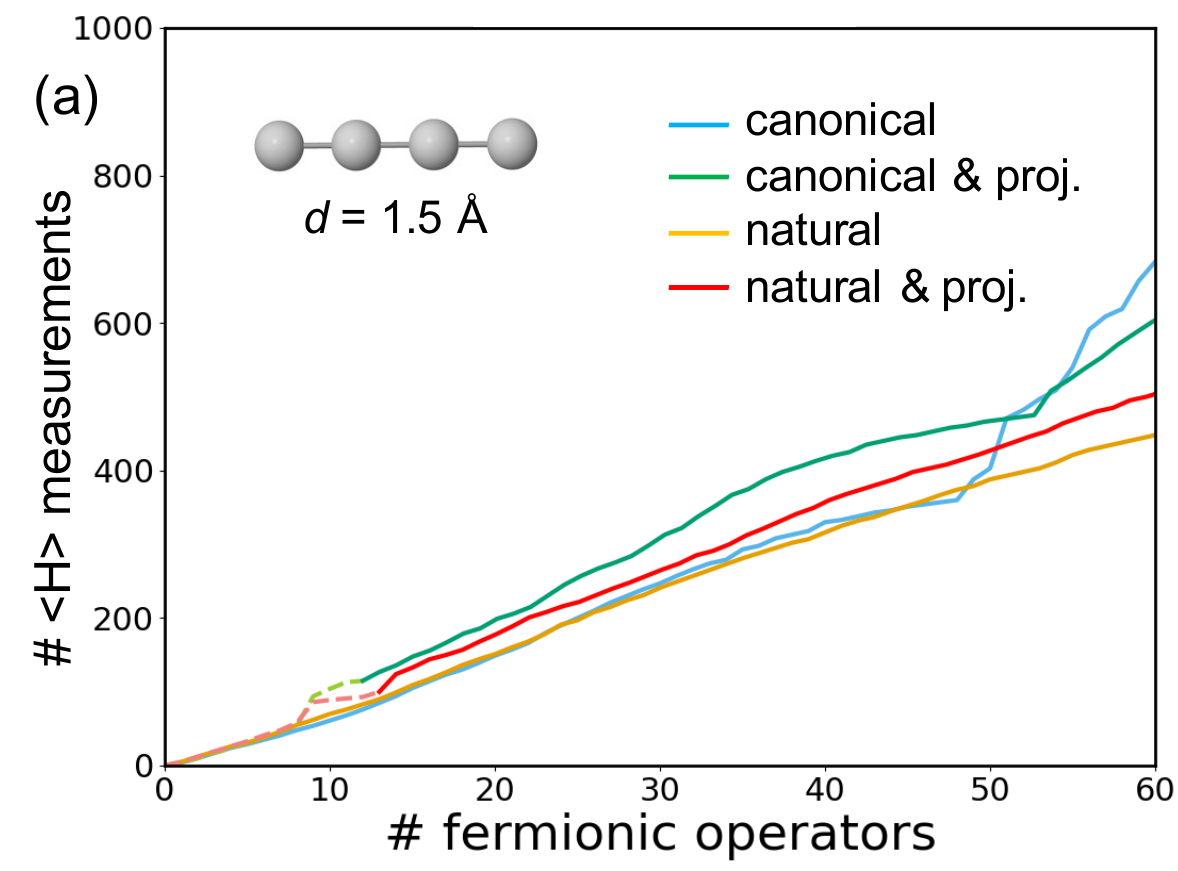}
    \includegraphics[width=7.3cm]{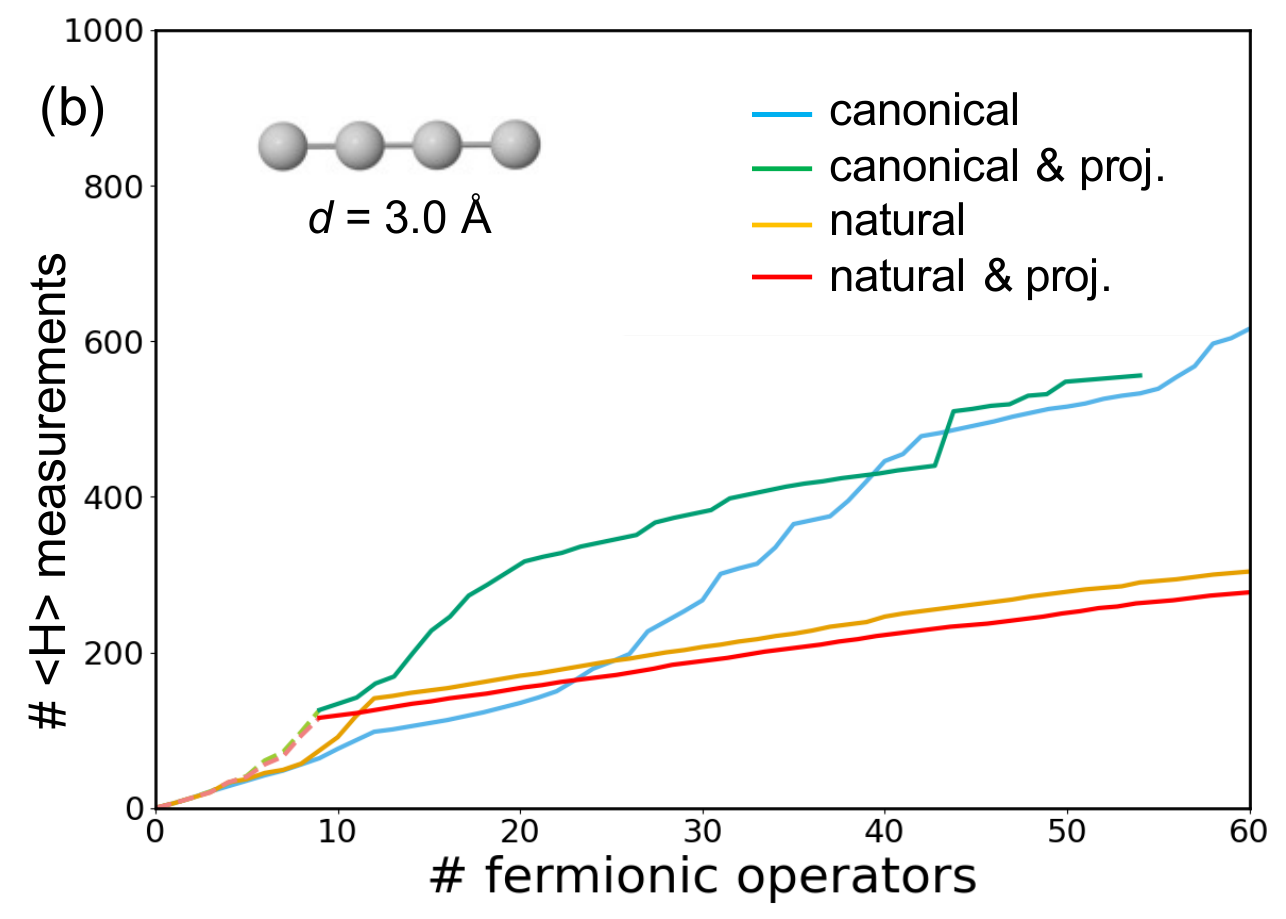}
    \caption{Number of energy measurements ($\langle H\rangle$) for the simulation of the linear \ce{H4} system with 1.5~{\AA} (a) and 3.0~{\AA} (b) \ce{H-H} distances performed with canonical (blue), natural (orange), projected canonical (green), and projected natural (red) orbitals.
    Initial iterations with orbital subspaces indicated with dashed lines.}
    \label{fig:h4_linear_its}
\end{figure}

We now turn our attention to the performance of the classical algorithm (BFGS) employed to optimize the set of ansatz's parameters $\{\theta_i\}$. 
Figure~\ref{fig:h4_linear_its} shows the cumulative count of energy evaluations requested by the classical optimizer throughout the ADAPT-VQE's procedure, plotted against the number of fermionic operators.
This analysis is conducted for the linear \ce{H4} system, employing the various approaches described in the preceding sections.

In general, the cumulative count of energy evaluations exhibits a nearly linear increase with the number of fermionic operators. 
This suggests that the additional energy measurements requested by the classical optimizer for each new operator remain roughly constant. 
However, notable changes in tendency occasionally occur at specific numbers of operators, which could be associated with significant alterations in the wavefunction as determined by parameters $\{\theta_i\}$. 
These changes in the growth of the number of energy measurements align with variations in the profile of the energy errors (Figure~\ref{fig:h4_linear_orbs}). 
Notable instances of this relationship are observed for $d=1.5$~{\AA} using canonical MOs and NOs with 50 and 75 operators, respectively, and for $d=3.0$~{\AA} at 10 operators.

We observe that NOs generally exhibit a smaller slope, indicative of a potential reduction in the computational cost.
This difference is particularly pronounced for $d=3.0$~{\AA}, where the slopes obtained using canonical MOs and NOs are notably distinct. 
As previously observed in our analyses, the advantage of NOs is maximized in strongly correlated geometries.
On the other hand, employing active space projection schemes, despite the reduction in the number of qubits, does not notably impact the number of necessary energy evaluations, irrespective of whether canonical MOs or NOs are utilized.

In addition to measuring the Hamiltonian, ADAPT-VQE necessitates the quantum computer to compute the energy gradients concerning each operator in the pool. Since the quantity of operators remains constant throughout the iterative process, the number of gradient evaluations remains constant as well. Clearly, incorporating NOs does not alter the operator count, thus maintaining the same number of gradients as in calculations employing canonical orbitals.
Conversely, the projection technique entails a reduction in the space of accessible operators within the subsystem, thereby necessitating fewer gradient evaluations.

\subsection{Application to the \ce{H2O} molecule} \label{sec:h2o}

\begin{figure}[H]
    \centering
    \includegraphics[width=6.8cm]{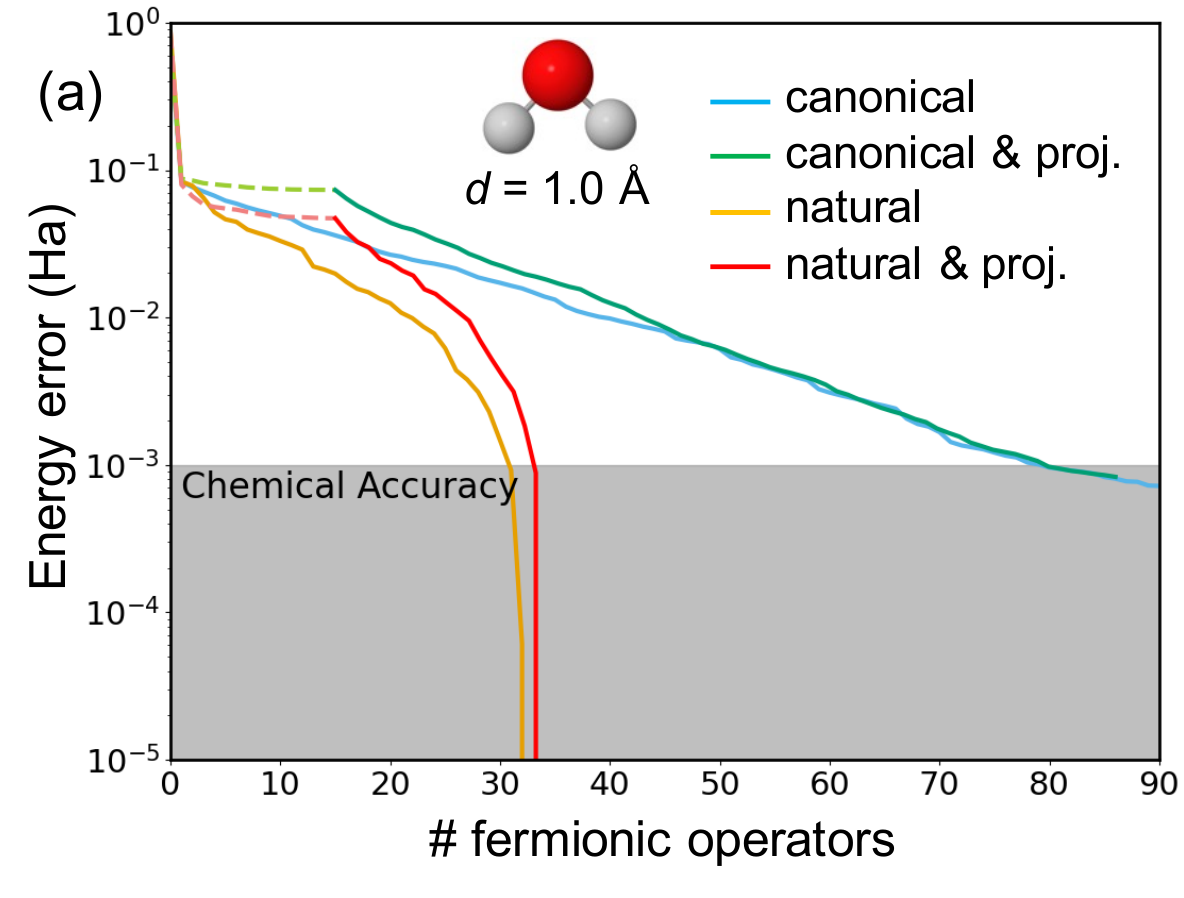}
    \includegraphics[width=7.1cm]{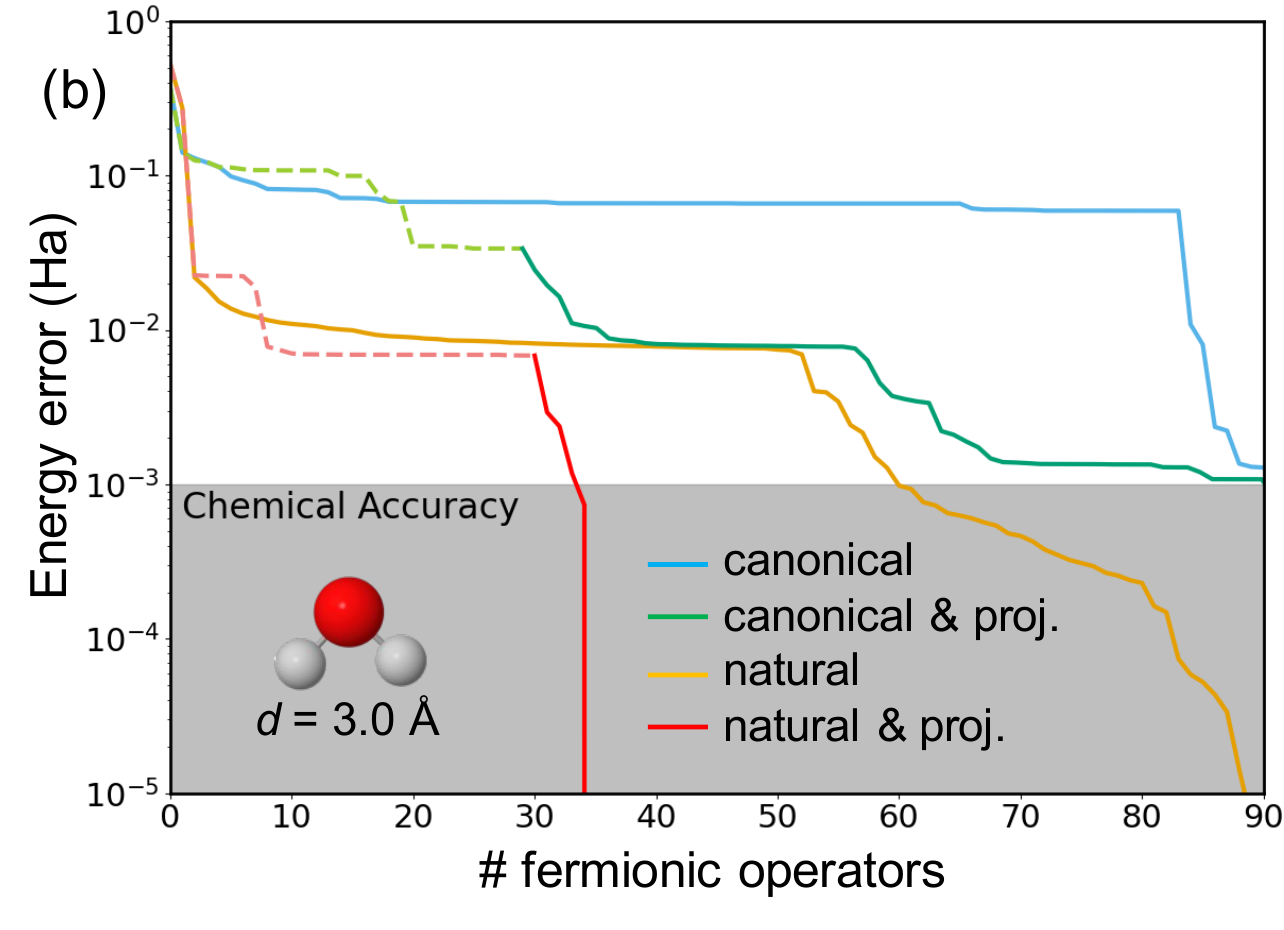}
    \caption{Energy errors (in Hartree) for the simulation of the water molecule with bond length of 1~{\AA} (a) and 3~{\AA} (b) obtained with canonical (blue), natural (orange), projected canonical (green), and projected natural (red) orbitals. 
    Initial iterations with orbital subspaces indicated with dashed lines.}
    \label{fig:h20_energy_d1}
\end{figure}

In this section, we evaluate the applicability of these strategies to a larger system, that is, the water molecule.
As in the previous sections, we compare standard ADAPT-VQE results with those obtained with the two strategies proposed in this paper.
Figure~\ref{fig:h20_energy_d1} presents energy errors as a function of the number of fermionic operators with respect to CAS(8,10).

At the near-equilibrium structure (\ce{OH} bond length of 1~{\AA}), the decay of energy errors within the initial ADAPT steps is similar with all approaches.
But, after a small number of added operators, calculations with NOs outperform those with the canonical basis, reaching chemical accuracy with a relatively compact wavefunction ($\sim30$ operators).  
It is also worth noticing that the basis projection scheme with the use of NOs yields basically the same results than those considering all NOs from the first iteration. 
Hence, in this case, the basis projection strategy does not exhibit any improvement in terms of energy accuracy, but still provides the advantage of using fewer qubits for a segment of the simulation. 

Converging towards chemical accuracy in the stretched molecule poses a significantly challenging task. 
Employing canonical orbitals fails to achieve chemical accuracy even after 80 iterations (fermionic operators in the ansatz), leading to energy errors on the order of 0.1 Hartrees. 
This limitation is significantly alleviated by the projection of the subspace ADAPT-VQE solution, which effectively guides the growth of the wavefunction, reaching chemical accuracy after 90 steps. 
In contrast, leveraging NOs markedly enhances the performance of the ansatz. 
The positive impact of the NO basis is particularly pronounced within the initial operators, where the energy error decreases much more rapidly than with the canonical basis.
Interestingly, the projection scheme with the NO basis efficiently steers the ansatz, resulting in much faster convergence of energies compared to its non-projected counterpart.

\begin{figure}[H]
    \centering
    \includegraphics[width=6.8cm]{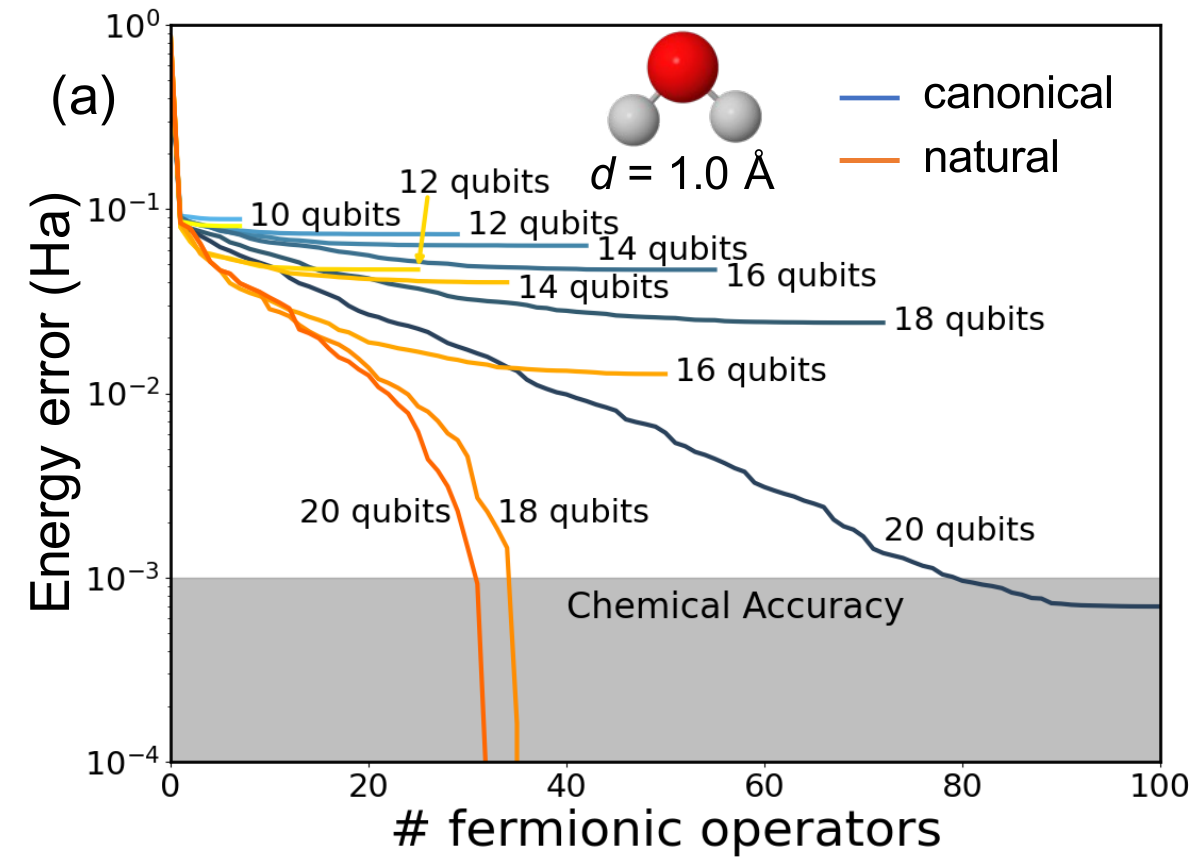}
    \includegraphics[width=7.2cm]{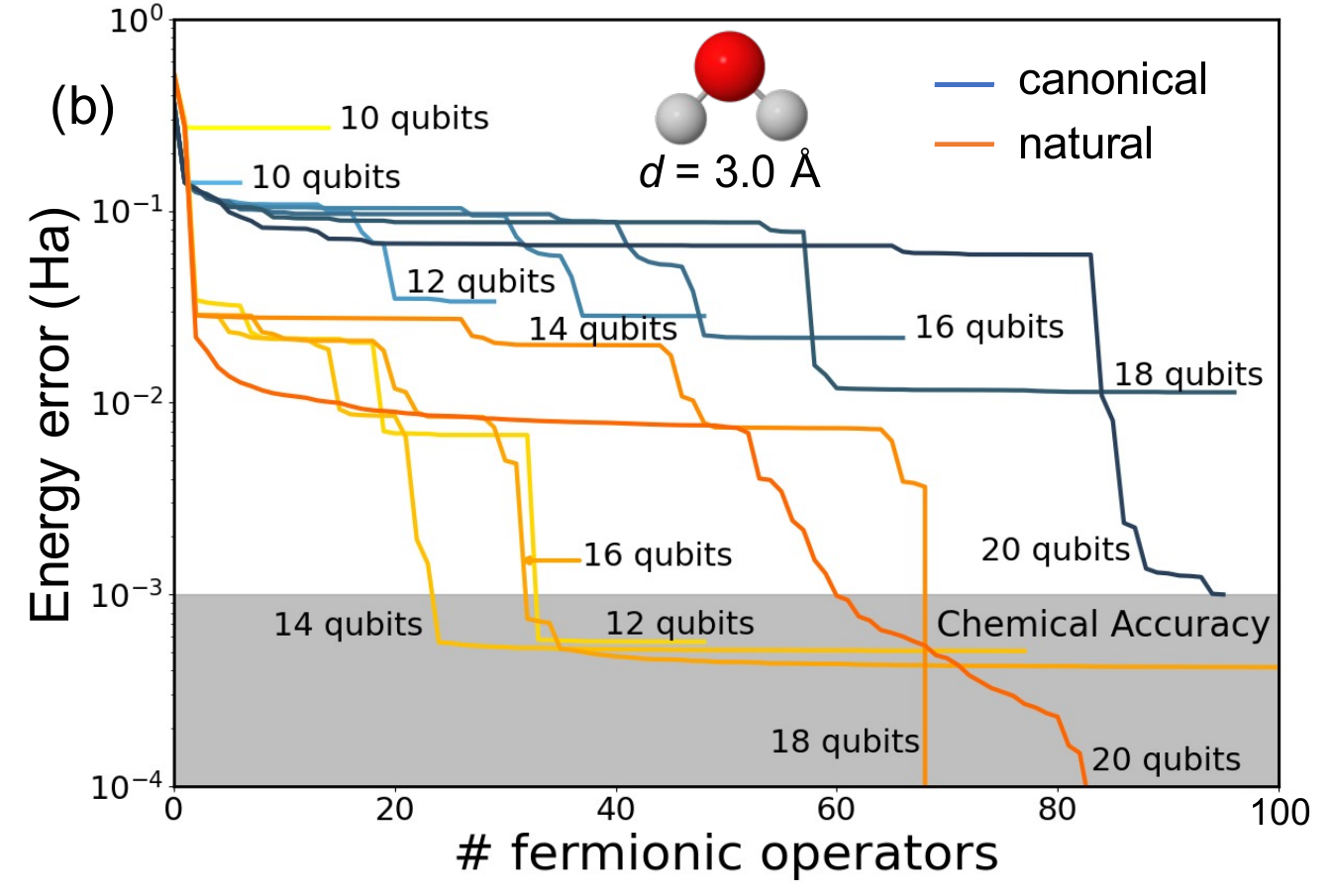}
    \caption{Energy errors (in Hartree) for the simulation of the \ce{H2O} molecule with 1~{\AA} (a) and 3.0~{\AA} (b) \ce{O-H} bond distances obtained with canonical (blue) and natural (orange) orbitals, and considering different number of qubits (active orbitals).}
    \label{fig:h20_energy_1_5}
\end{figure}

Next, we analyze the dependence of the energy errors with the number of available orbitals (or qubits) for the two geometries.
Figure~\ref{fig:h20_energy_1_5} illustrates the results for active spaces ranging from 5 to 10 orbitals (equivalent to 10 to 20 qubits).

In summary, the computed energy profiles for both near-equilibrium and elongated bonds underscore the superiority of NOs in approximating energies close to the exact solution.
Furthermore, similar to the observations in the \ce{H4} models, calculations involving reduced orbital spaces in open-shell structures (stretched bonds) manifest clear improvements, suggesting a more facile exploration of optimal paths for constructing the ground state wavefunction. 
These findings provide further validation for the utility of the NO basis and the projection scheme in characterizing strongly correlated systems.

\subsection{Improvements in the circuit depth} \label{sec:depth}

One critical parameter defining a quantum circuit is its depth, which signifies the maximum count of gates, or qubit unitary transformations, along any pathway within the circuit. 
For adaptive algorithms to effectively operate on NISQ devices, it becomes absolutely necessary to minimize circuit depths, as (ideally) all gates in the circuit should be executed within the coherence time limit of the qubits. 
Circuits with increased depth correspond to elevated noise levels, thereby requiring a greater number of samples to accurately measure Hamiltonian expectation values.
In the following, we quantitatively assess the impact of the introduced strategies on the depth of ADAPT-VQE circuits. 
These results are obtained through simulations involving the implementation of quantum circuits using the staircase algorithm\cite{nielsen2010quantum} in combination with the Trotterization of each linear combination of qubit operators representing a fermionic excitation, 
while considering statistical errors in the energy measurement.
It's important to note that in these simulations we do not account for noise effects.

\begin{figure}[H]
    \centering
    \includegraphics[width=6.8cm]{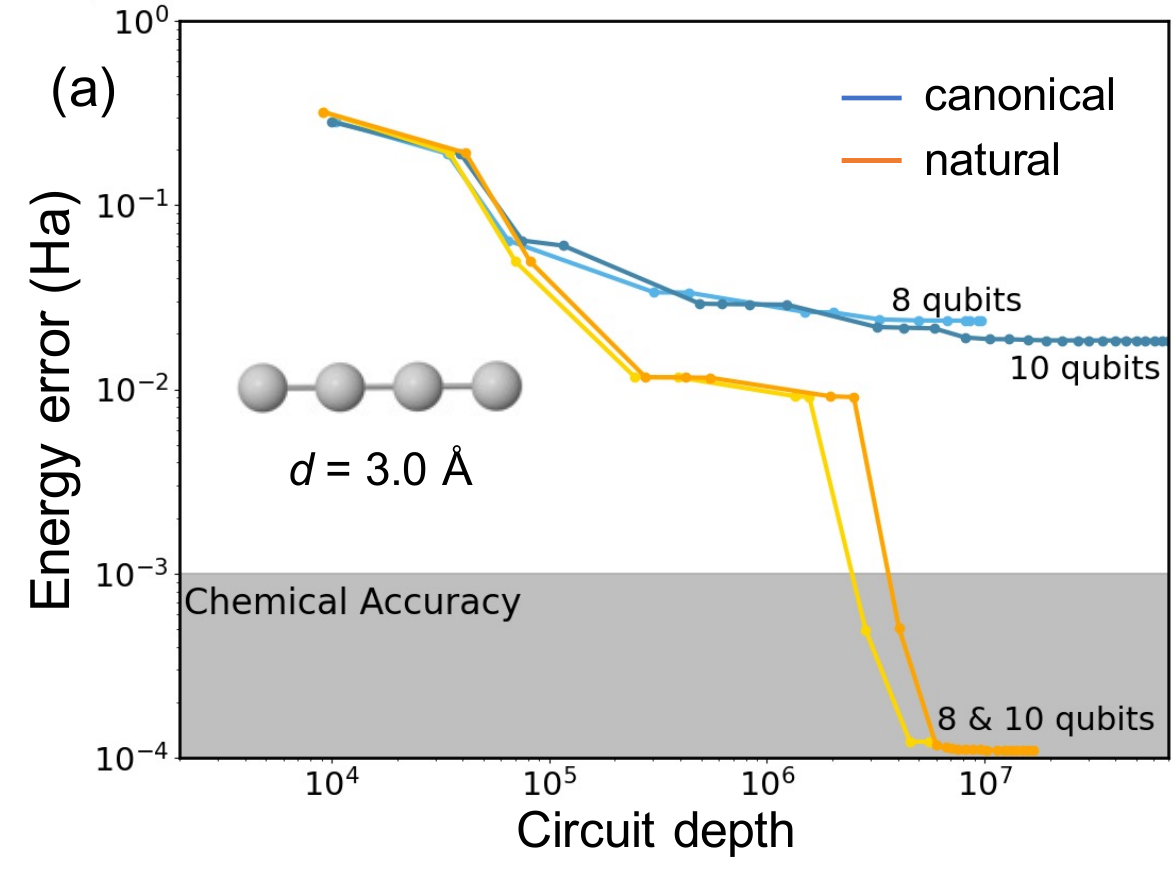}
    \includegraphics[width=6.8cm]{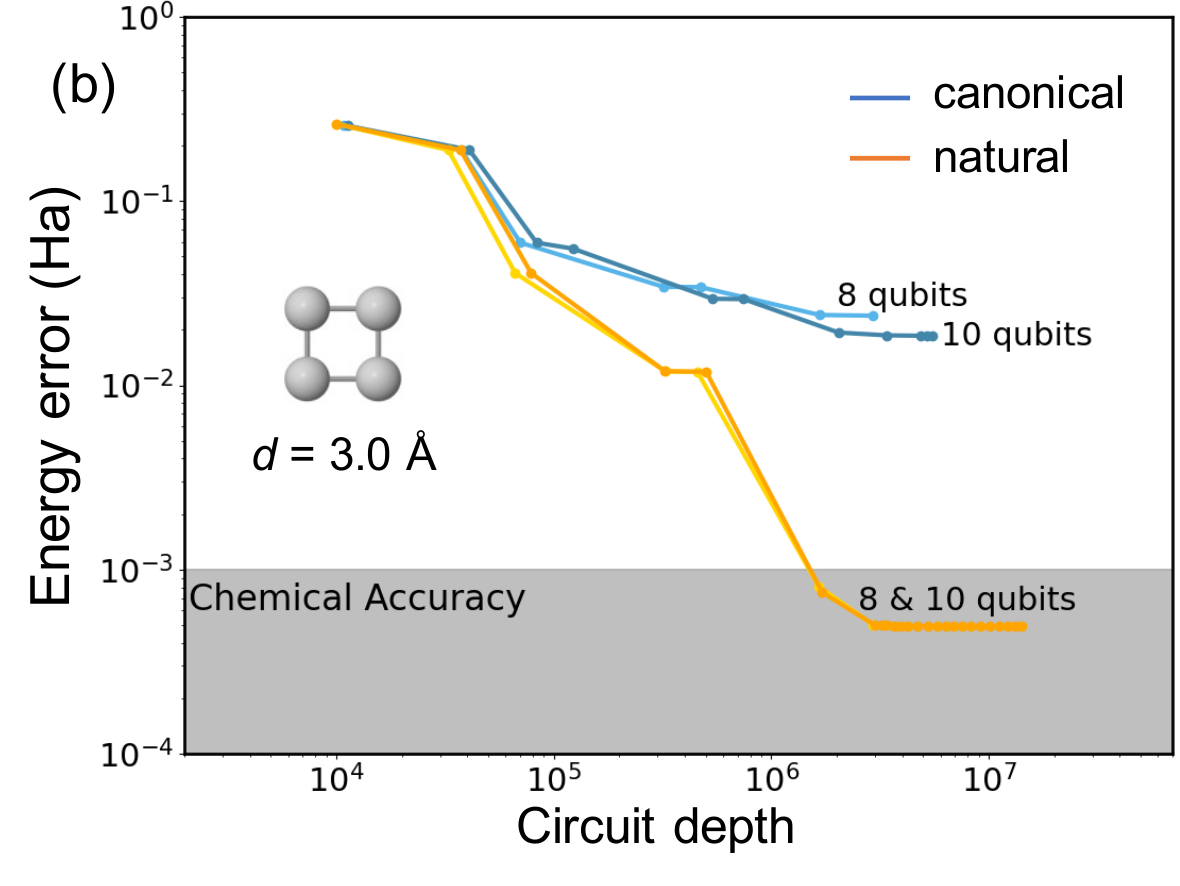}
    \includegraphics[width=6.8cm]{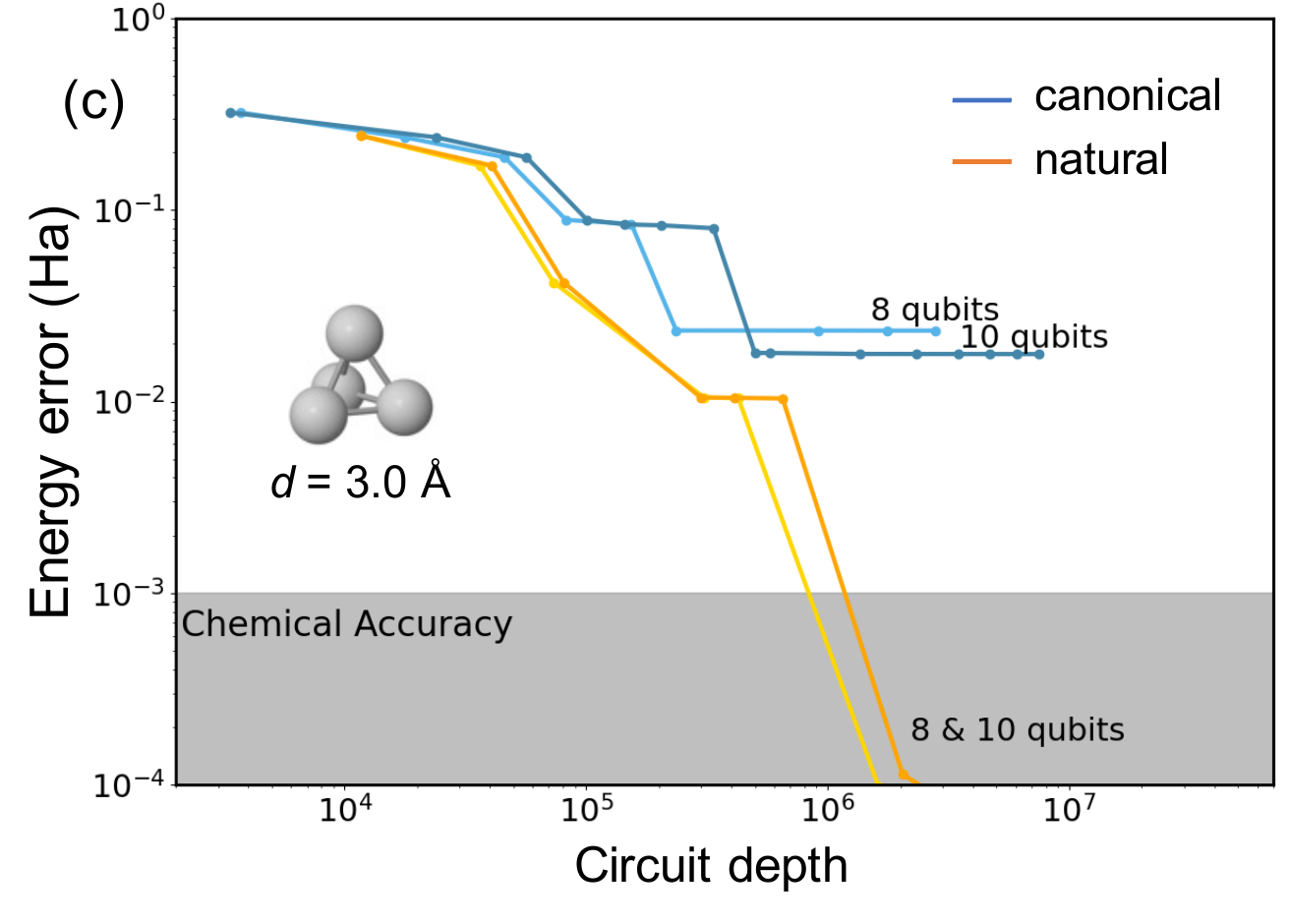}
    \caption{Energy errors (in Hartree) as a function of the cumulative total circuit depth calculated for the linear (a), square (b) and tetrahedral (c) \ce{H4} models with $d=3.0$~\AA, using canonical (blue) and natural (orange) orbitals, and with two active spaces (4 and 5 orbitals).}
    \label{fig:trotter_h4all}
\end{figure}

Figure~\ref{fig:trotter_h4all} shows the increase in the cumulative total circuit depth as the size of the ADAPT-VQE ansatz grows for the linear, square and tetrahedral \ce{H4} systems at the stretched geometries ($d = 3$~\AA). 
Here, we define the total circuit depth as the sum of the depth of all the evaluated circuits for all iterations. We use this as an estimated measure of the real cost of the algorithm.
In all three arrangements, the utilization of NOs substantially diminishes the circuit depth necessary to reach chemical accuracy, inline with the dependence of energy errors with the number of fermionic operators. 
Notice that the total depth of the circuits directly correlate with the total number of two-qubit controlled NOT (CNOT) gates, holding higher error rates than single-quibit gates. 
Indeed, representation of energy errors with respect the number of CNOTs follow identical profiles than those in Figure~\ref{fig:trotter_h4all} (Figure~\ref{si:fig:trotter_h4all}). 
We observed that the depth of individual circuits used to encode the electronic states increases roughly linearly with respect to the number of operators added to the ansatz. In the studied \ce{H4} systems, the constant of proportionality is around 100-200 depth per operator.  

\section*{Conclusions}

In summary, our research introduces and investigates two straightforward yet impactful strategies to enhance the performance of ADAPT-VQE ans\"atze. Firstly, we address the crucial aspect of initial state preparation by leveraging the NOs derived from the UHF density. Secondly, we target the growth of the wavefunction by incorporating subspace orbital solutions as intermediates, projecting them onto the complete orbital space.
Finally, we have also investigated the synergies emerging from combining both ideas.
The application of these simple approaches to compute electronic energies for \ce{H4} arrangements and the \ce{H2O} molecule, featuring diverse geometries, has yielded remarkably positive results. Notably, substantial enhancements have been observed in comparison to the canonical ADAPT-VQE. Specifically, both strategies yield more compact ans\"atze, particularly excelling in characterizing open-shell scenarios, such as those encountered in \ce{H4} systems with significant interatomic distances and the stretched water molecule.
These outcomes underscore the efficacy of NOs, especially in achieving chemical accuracy with concise ans\"atze, a critical attribute for unlocking the potential of quantum computing in electronic structure simulations. The significant advantages of NOs become apparent when contrasted with the inherent limitations of canonical orbitals when dealing with compact wavefunctions.
The reduced computational cost associated with NOs, especially in challenging scenarios, can be ascribed to their intrinsic multiconfigurational nature. 
This feature facilitates a more rapid convergence of the wavefunction toward the desired solution, consequently diminishing the number of required energy measurements for classical optimization and the depth of the implemented circuits. 
This efficiency is particularly valuable in the context of NISQ devices, where resource constraints demand a delicate balance between accuracy and computational cost.
In light of our findings, we advocate for the systematic integration of NOs and/or projection techniques in ADAPT-VQE, particularly for the investigation of strongly correlated (open-shell) molecules. 
We firmly believe that the present results provide a compelling justification for the adoption of these strategies, paving the way for more efficient and accurate quantum simulations in electronic structure studies.
Moreover, these physically-inspired strategies could be combined with hardware efficient schemes, where more hardware efficient pools are used\cite{Tang:adapt:2021,Yordanov:adapt:2021} and where the circuits are constructed more densely, as in TETRIS-ADAPT-VQE.\cite{Anastasiou:tetris-adapt:2024}
Finally, we emphasize that, in this study, we have overlooked the significant noise effects inherent in NISQ quantum computers. 
We intend to explore the impact of noise on various forms of ADAPT-VQE in future investigations.

\begin{acknowledgement}
This work was funded by the Spanish Government MICINN (project PID2022-136231NB-I00), the Gipuzkoa Provincial Council (project QUAN-000021-01), the European Union (project NextGenerationEU/PRTR-C17.I1), as well as by the IKUR Strategy under the collaboration agreement between Ikerbasque Foundation and DIPC on behalf of the Department of Education of the Basque Government.
The authors acknowledge the financial support received from the BasQ Strategy under the collaboration agreement between Ikerbasque Foundation and DIPC on behalf of the Department of Education of the Basque Government.
The authors are thankful for the technical and human support provided by the DIPC Computer Center. 
D.C. is thankful for financial support from IKERBASQUE (Basque Foundation for Science). 
\end{acknowledgement}

\begin{suppinfo}
Orbital symmetry analysis.
Additional results on energy errors, fidelity of the density matrix, energy measurements, and number of CNOT gates in the generated circuits.
\end{suppinfo}

\section*{Data Availability Statement}
The data that supports the findings of this study are available within the article and its supplementary material, which contains additional and complementary data for the studied systems.

\section*{Conflict of Interest} 
The authors declare that they have no conflict of interest.

\bibliography{abbr,references,active_space} 

\end{document}